%% file: galMagRM.tex
\documentclass[12pt, twocolumn, trackchanges, ]{aastex631} 
\usepackage{amsmath} 
\usepackage{graphicx}
\usepackage{xcolor}
\usepackage{soul}

\newcommand{\joeltext}[1]{{\color{blue} #1}}

\usepackage{tipa}            
\newif\ifAMStwofonts
\AMStwofontstrue
\usepackage{ulem}          
\usepackage{mathtools}  
\usepackage{gensymb}

\def\arcdeg{\hbox{$^\circ$}}
\def\degree{$^{\circ}$}
\def\ga{\mathrel{\hbox{\rlap{\hbox{\lower4pt\hbox{$\sim$}}}\hbox{$>$}}}}
\def\la{\mathrel{\hbox{\rlap{\hbox{\lower4pt\hbox{$\sim$}}}\hbox{$<$}}}}

\accepted{\apj}

\begin{document}
\title{Determining the Magnetic Field in the Galactic Plane from New Arecibo Pulsar Faraday Rotation Measurements}
\shorttitle{Determining the Magnetic Field in the Galactic Plane}

\shortauthors{Curtin,  Weisberg, \& Rankin}

\input{auth.tex}

\correspondingauthor{Alice P. Curtin}
\email{alice.curtin@mail.mcgill.ca}

\begin{abstract}
We develop a new method for studying the Galactic magnetic field along the spiral arms using pulsar Faraday rotation measures (RMs). Our new technique accounts for the dot-product nature of Faraday rotation and also splits the associated path integral into segments corresponding to particular zones along the LOS. We apply this geometrically-corrected, arm-by-arm technique to the low-latitude portion of a recently published set of Arecibo Faraday RMs for 313  pulsars along with previously obtained RMs in the same regions. We find disparities $>1\sigma$ between the magnitude of the field above and below the plane in the Local Arm, Sagittarius Arm, Sagittarius-to-Scutum Interarm, the Scutum Arm, and Perseus Arm. We find evidence for a single field reversal near the Local/Sagittarius arm boundary. Interestingly, our results suggest that this field reversal is dependent on latitude, occurring inside the Sagittarius arm at negative Galactic latitudes and at the Local-to-Sagittarius Arm boundary at positive Galactic latitudes.
We discuss all of our results in the context of different models and other observational Galactic magnetic field analyses. 
\end{abstract}


\section{Introduction}

Pulsars are excellent probes of the  magnetic field in the plane of the Galaxy due to the phenomena of Faraday rotation and dispersion.  A radio source's Faraday rotation measure (RM), measured in rad m$^{-2}$, can be determined via observations of the rotation of its linear plane of polarization as a function of frequency, and is related to the vector magnetic field, sampled along the line of sight (hereafter LOS), as follows:

\begin{equation}
{\rm{RM}}=\frac{e^3}{2\pi m_e^2 c^4}\int_{\rm source}^{\Earth} n_e {\bf{B}} \cdot d{\bf{s} }   ,
\label{eqn:RMdef}
\end{equation}
where $e$ and $m_e$ are the electron charge and mass, c is the speed of light, $n_e$ is the free electron density, \rm {\bf B} is the magnetic field, and d{\bf{s}} is an element of path length along the source-observer LOS \citep{LK2012}.  

A pulsar's dispersion measure (DM), measured in pc~cm$^{-3}$, can be determined from observations of the pulse delay as a function of frequency, and is related to the electron density along the LOS through
 \citep{LK2012} :

\begin{equation}
{\rm{DM}}= \int_{\rm source}^{\Earth} n_e\  ds.
\label{eqn:DMdef}
\end{equation}
 
Specifically, Eqs.~(\ref{eqn:RMdef}) and~(\ref{eqn:DMdef}) indicate that a pulsar's  Faraday-rotated, 
dispersed signals reveal the mean of the magnetic field magnitude $|B|$ multiplied by a cosine term originating from the dot-product nature of Eq.~(\ref{eqn:RMdef}):
\begin{equation}
\langle |B| \cos \theta_{\scriptsize{\texttoptiebar{B\ ds}}} \rangle = 1.232\ \frac{\rm{RM}}{ \rm{DM} }\  \mu\rm{G},
\label{eqn:BcosTheta}
\end{equation}
when  RM and DM are measured in their usual units, rad m$^{-2}$ and pc cm$^{-3}$, respectively, and where $\theta_{\scriptsize{\texttoptiebar{B\ ds}}} $, is the angle between the source-Earth LOS and magnetic field vectors.
The angular brackets represent an average along the path over which RM and DM are evaluated. One can determine the mean magnetic field magnitude {\it{alone}} from Eq. (\ref{eqn:BcosTheta}) if the geometry of the field (and hence the value of the cosine term) are known.

Polarized extragalactic radio sources (EGS)
allow measurement of the total RM, but this only reveals the {\it{integral}} of the rather complicated quantity 
$(n_{\rm e} \ |B| \  ds \ \cos \theta_{\scriptsize{\texttoptiebar{B\ ds}}})$ along the full path from the source to the observer. This impedes any effort to disentangle 
the magnetic field  from the other variables, especially in a particular galactic zone.
Pulsed sources such as pulsars and fast radio bursts are the only class of object for which a DM can be measured, and hence the only objects to which Equation (\ref{eqn:BcosTheta}) can be directly applied.  

Furthermore, because pulsars are embedded 
within the interstellar 
medium of our Galaxy at approximately known distances, we can use groups of them to assess the systematic galactic magnetic field along specific {\it{intragalactic}} zones such as spiral arms or interarm regions, including regions lying beyond other systematic magnetic zones.  In this work, we present a new ``arm-by-arm'' technique for determining the systematic magnetic field along long stretches of spiral arms by correcting for the assumed field geometry and for the effects of nearer magnetic regions. 

In a companion paper, \citet{Rankin2023} have taken advantage of the sensitivity of the Arecibo telescope and 
its wide-bandwidth Gregorian feed and spectrometers to measure RMs of virtually all known low-latitude 
pulsars in the telescope's accessible declination range, in addition to some higher-latitude sources. We use these new low-latitude measurements and all other relevant pulsar RM determinations in and near
these directions to study the systematic Galactic magnetic field in these portions of the Galactic plane. We believe that our work represents the first geometrically-corrected 
measurement of the systematic magnetic field magnitude along large stretches of spiral arms and interarm regions. 
 
The plan of this paper is as follows: In \S\ref{sssec:data}, we discuss the measurements used in our Galactic magnetic field analysis. 
In \S\ref{sssec:analysis}, we first present our new ``arm-by-arm'' technique for studying the Galactic magnetic field. Then, in \S\ref{sssec:systematic}, we apply our  technique to study the systematic field in the plane in and near our observed Galactic longitude ranges. In \S\ref{sssec:Discuss}, we discuss our results within the larger context of Galactic field models. A summary is presented in \S\ref{sssec:Concl}.

\section{Pulsar Rotation Measures, Distances, and Galactic Structure}
\label{sssec:data}

\subsection{ Rotation Measures}
\citet{Rankin2023} have determined the RMs of 313 pulsars within the field of view of Arecibo Observatory in the longitude ranges $\sim$\! (30$\degr - 80\degr$) and~$\sim$\! (175$\degr - 210\degr$). Their observations were carried out at frequencies near 0.3 and 1.5 GHz, beginning in early 2003 and concluding in late  2020. RMs  and uncertainties for this set of pulsars can be found in Table 1 of \citet{Rankin2023}. Fig. \ref{fig:AreciboFace} displays the Galactic locations of the pulsars from that work that we include in our analysis. 

For the current analyses, the low-latitude ($|b| \leq 9\degr$) portion of these new RMs are supplemented by previously measured ones in the Galactic plane in nearby directions, taken from  Version 1.64 of  the ATNF Pulsar Catalogue \citep{Manch2005}\footnote{Some of the RMs  reported in the ATNF Catalogue include an unsubtracted ionospheric component, which indicates that some of the ATNF Catalogue's RMs have larger uncertainties than quoted.} plus  a set of more recent observations {\citep{Johnston2021}.} In the case of multiple RM values for a particular pulsar, we adopted that with the smallest uncertainty. While all pulsars studied will have $|b| \leq 9\degr$, the term ``low-latitude'' will henceforth be omitted in most cases.

In addition to pulsars, we could also use the RMs from polarized EGS to study the Galactic magnetic field. However, as our method (detailed in \S\ref{sssec:analysis}) explicitly depends upon separating the LOS from a source into known segments, we do not rely on RMs from EGS, except when commenting on others' work.

\subsection{Additional Information }


\subsubsection{Distances } \label{sssec:dist}

For almost all of the pulsars' LOS, we adopted the ``best'' pulsar distance from Version 1.64 of the ATNF Pulsar Catalogue\footnote{This is denoted as \textsc{``Dist''} in that Catalogue.}. This distance is defined as  the distance 
derived from the  measured DM and the \citet{YMW2017} [hereafter YMW17] electron density model, unless a more accurate distance is 
available from  other techniques such as VLBI parallax measurement, kinematic distance via HI 21 cm absorption, or
association with another object of known distance.\footnote{The single exception to these rules was PSR J2021+3651, where we  replaced the ATNF  Catalogue's \textsc{``Dist''}  value of 1.80 kpc with the YMW17 DM distance estimate of 10.51 kpc. This pulsar has a very high DM  for its putative \textsc{``Dist,''} which was derived from X-ray measurements of the  hydrogen  column density along the LOS \citep{Kiriet2015}. Nevertheless, \citet{Kutkcu2022} also suggest a relatively high distance similar to ours, based upon analogous considerations. While a localized  electron density enhancement could explain the disproportionately high DM, there is no evidence of an HII region \citep{Ocker2024} along the LOS.  Similarly, we find no unusual HI column density \citep{HI4PIpaper} along the LOS. We believe that it is more likely that the X-ray distance determination procedure, which is less well-established than the DM distance technique, has rendered an unrealistically small distance.   We note this one exception in Table 1 of \citet{Rankin2023}.}

\subsubsection{Galactic Structure}
 \label{sssec:galstruct}
\begin{figure}[t]
\includegraphics[trim=1.1in   5in 0in  1.1in, scale=0.70  ]{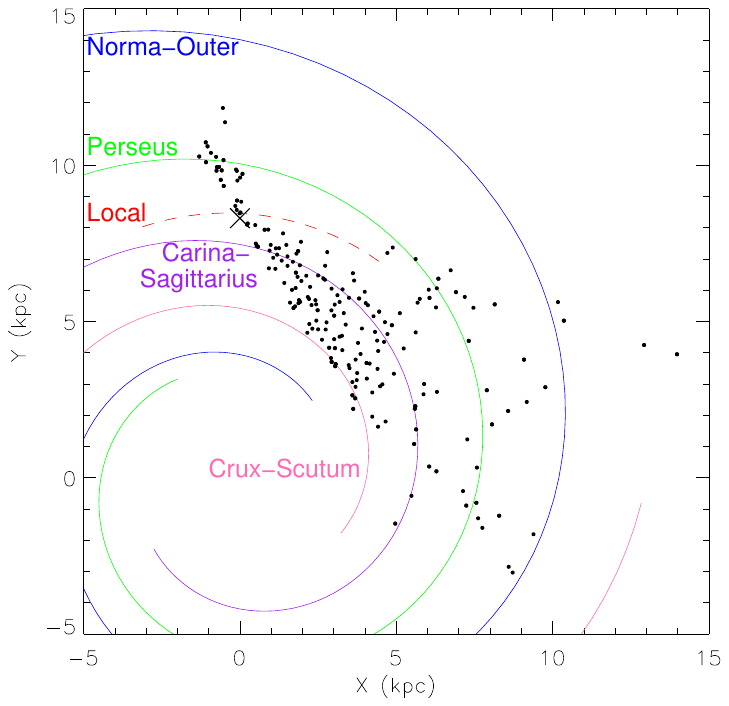} 
\caption{Locations of all low-latitude ($|b| \leq 9 \degr$) pulsars measured and tabulated in \citet{Rankin2023}. The Solar System is marked by the black $\times$ at ($x,\ y$) = (0.0,\ 8.3)\ kpc. The plotted locations of the spiral arm centers are from  YMW17 \citep{YMW2017}. 
} 
\label{fig:AreciboFace}
\end{figure}

One of the major goals of this work is to study the nature of the systematic magnetic field associated
with various Galactic structures, especially spiral arms and interarm regions.  Hence it is crucial 
to use the best available knowledge of these structures. With the exception of the Local Arm (see  
\S\ref{sssec:local}), we use the spiral arm centers 
as defined by \citet{HouHan2014},  and
slightly modified by YMW17, with the Solar System located 8.3 kpc from the Galactic center. The general logarithmic expression for the arm centers can be found in Eq. (10) of the latter reference, with the adopted values of constants therein  and Table 1 of  the latter reference. The arm centers can also all be seen in Fig. \ref{fig:AreciboFace}.

\section{Arm-by-arm Galactic Magnetic Field Analysis}
\label{sssec:analysis}

In this section, we develop a new technique for attributing portions of the path-integrated quantities RM and DM to specific
arms and interarm regions along each pulsar's LOS.  Once we determine the RM and DM attributable to each such zone,
we are able to also measure the $B$-field associated with it.  We call this procedure the ``arm-by-arm'' technique, since the procedure splits the LOS quantities outward from the Earth into contributions from each such zone.

Given our new RM values, we focus our magnetic field analyses on 
Arecibo-accessible spiral arm zones, interarm zones, and adjoining regions. In particular, we focus on parts of the Local, Carina-Sagittarius, Crux-Scutum, and Perseus Arms plus distinguishable regions between them. We searched these regions and their vicinities for systematic patterns of magnetic fields, guided partly by the results of the earlier \citet{Wetal2004} Arecibo RM survey, and also by subsequent  work. 
We do not comment on the magnetic field structure elsewhere in the Galaxy since we have no new information to contribute in those regions. Interested readers are directed to \citet{Han2018} and \citet{Xuet2022} for recent analyses of  other zones. Those authors and most others investigating the overall structure of the Galactic plane magnetic field conclude that it  lies parallel or antiparallel to spiral structures, often reversing between them and perhaps even between arm and interarm zones. However, the field direction (and magnitude) between reversals along with the number of reversals is still controversial. 

As a first approximation to the more complicated reality, we model the form of each arm
and relevant interarm region as  having a  constant magnetic 
field magnitude $|B|$ throughout, pointing either clockwise (CW, helicity $h=-1$) or counter-clockwise (CCW, helicity $h=+1$) 
along the arm or interarm region, bounded on its Galactocentric inner and outer edges by 
logarithmic spirals\joeltext{\footnote{Specifically, we model the field as being locally parallel to the
{\it{outer}} boundary in all cases.}}. For  our equation of Galactocentric radius $R$ as a function of 
Galactocentric azimuth $\phi$ for the $i^{th}$ logarithmic spiral inner or outer {\it{ boundary}}, we 
follow YMW17 Eq. (10):
\begin{equation}
\ln{\frac{R}{\ \ R_{0,i}} } = (\phi - \phi_{0,i}) \tan \psi_{i},
\label{eqn:spiral}
\end{equation}
where $R_{0,i}$ is the Galactocentric radius of the  $i^{th}$ spiral boundary at 
Galactocentric azimuth $\phi_{0,i}$, and 
$\psi_{i}$ is the  $i^{th}$ spiral pitch angle.

In concert with our zeroth-order assumption of constant field magnitude along the arm, we adjust the inner and outer edges of each such zone  from their nominal locations 
so as to maximize the consistency of the magnetic field within the 
zone. The parameters of the resulting  inner and outer boundaries for each zone 
are listed in Table \ref{table:Bs}, and those boundaries are delineated by dotted black curves in  face-on Galactic plots located in or near the section discussing each zone. 

\begin{deluxetable*}{lccrrrrrrccc}
\tablecolumns{11}
\tabletypesize{\footnotesize}
\tablecaption{Fitted Mean Magnetic Fields in Selected Regions of the Galactic Plane ($|b| \leq 9^{\rm o}$)  \label{table:Bs} }

\tablehead{
\colhead{Region} & \colhead{Section} &\multicolumn{6}{c}{Logarithmic spiral parameters for Eq. (\ref{eqn:spiral})    }  &
&\multicolumn{2}{c}{Mean magnetic field\tablenotemark{a} }   \\
\cline{3-8} \cline{10-11}
 \colhead{  } &  & \colhead{$R_{0,\rm{inner}}$ }   &   \colhead{$\phi_{0,\rm{inner}}$ }  &  \colhead{$\psi_{\rm{inner}}$}  & \colhead{$R_{0,\rm{outer}}$ }   &  \colhead{$\phi_{0,\rm{outer}}$}  &  \colhead{$\psi_{\rm{outer}}$}  &    &
  \colhead{For $-9^{\rm o} \leq b < 0^{\rm o}$} & \colhead{For $0^{\rm o} \leq b \leq 9^{\rm o}$}\\
   & &  \colhead{(kpc)}        &  \colhead{($^{\rm o}$)}  & \colhead{($^{\rm o}$)}  & \colhead{(kpc)} &  \colhead{($^{\rm o}$)} &\colhead{($^{\rm o})$}  &  & \colhead{($\mu$G)}& \colhead{($\mu$G)}\\
}
\startdata
Local Arm                         & \S\ref{sssec:local}         &  3.880 & 218.3  & 10.38   & 3.407 & 119.9  &  9.84 &  & -2.48 $\pm$ 0.42 & \phn-1.29       $\pm$ 0.62  \\  
Sagittarius Arm                   & \S\ref{sssec:Sag}           &  3.190 & 218.3  & 10.38   & 3.880 & 218.3  & 10.38 &  & \phn0.32 $\pm$ 0.32 & \phn\phn1.88 $\pm$ 0.24  \\  
Sag. to Scut. Interarm    & \S\ref{sssec:interSag2Scut} & 2.760  & 218.3  & 10.38   & 3.190 & 218.3  & 10.38 &  & \phn1.00 $\pm$ 0.29 & \phn-0.25    $\pm$ 0.36  \\
Scutum Arm                        & \S\ref{sssec:Scutum}        & 3.520  & 330.3  & 10.54   & 2.760 & 218.3  & 10.38 &  & \phn2.06 $\pm$ 0.36 & \phn\phn1.00 $\pm$ 0.26  \\
Perseus Arm                       &   \S\ref{sssec:Per}         & 3.407 & 119.9   & 9.84    & 3.807 & 119.9  &  9.84 &  &  -1.69 $\pm$ 0.70 & \phn-2.11    $\pm$ 0.55  \\
\enddata 
\tablenotetext{a}{A negative (positive) sign, synonymous with $h_k$= -1 (+1) in Eq. \ref{eqn:Bmagmeanandhelicity}, indicates a 
                   CW (CCW) magnetic field, as seen from North 
Galactic Pole. In all cases,  the field is modeled as pointing parallel to the outer spiral boundary of the region. } 
\end{deluxetable*}

Given the path-integral nature of the basic observables RM$_{\rm PSR}$ and DM$_{\rm PSR}$, it is useful
to break the {\it{total}} paths of Eqs.~(\ref{eqn:RMdef}) and~(\ref{eqn:DMdef}) into convenient subpaths,  
corresponding to the  $k$ distinguishable arms and  interarm regions  along the LOS from the $j^{th}$ pulsar:  

\begin{equation}
\rm{RM_{PSR\ j}} =    \sum_{i=1}^{k} \delta RM_{\rm i,j} \ \ \rm{rad \ m}^{-2},
\label{eqn:RMbroke}
\end{equation}
and
\begin{equation}
\rm{DM_{PSR\ j}} =    \sum_{i=1}^{k} \delta DM_{\rm i,j} \ \ \rm{pc \ cm}^{-3}.
\label{eqn:DMbroke}
\end{equation}

Since the LOS from one zone may pierce other zones, we will need to adequately model
the magnetic and electron-density properties of the  $i=1, 2, 3, ..., k-1$ intervening zone(s) in order to isolate the desired 
properties in the target, $k^{th}$, zone.  This principle also
dictates the order of analysis of the zones. Therefore,
in the above two equations, the first subpath corresponds to the Local Arm within which we are embedded,  higher-numbered ones represent successively farther zones along
the LOS from Earth, and the final, $k^{th}$ term corresponds to the farthest subpath, which is  bounded by the  pulsar and the earthward
edge of the zone  within which it lies.\footnote{The outward (away from Earth) ordering of the summation terms in Eqs. (\ref{eqn:RMbroke}) and 
(\ref{eqn:DMbroke}) always associates $i=1$ with the Local Arm; and is opposite to the inward (earthward) 
direction of the RM and DM path element {\it{d}}s and path integrals  of Eqs. (\ref{eqn:RMdef}) and 
(\ref{eqn:DMdef}) and any associated subpaths. Nevertheless, the sign of each 
summation term is set by the result of the associated path integral.}  This latter term, the contribution 
of the farthest zone, can then be
 isolated by subtracting the already-modeled contributions of all $k-1$ 
 intervening zones along the $j^{th}$ pulsar's LOS (unless the pulsar
 lies in the Local Arm, in which case there are {\it{no}} intervening zones and $k \equiv 1$):  
 
\begin{equation}
\delta \rm{RM}_{\rm j,k} = \rm{RM_{PSR, \ j}}  -    \sum_{i=1}^{k-1} \delta RM_{\rm i,j} \ \ \rm{rad \ m}^{-2},
\label{eqn:deltaRMbroke}
\end{equation}
and
\begin{equation}
\delta \rm{DM}_{\rm j,k} = \rm{DM_{PSR,\ j}}  -    \sum_{i=1}^{k-1} \delta DM_{\rm i,j} \ \ \rm{pc \ cm}^{-3} ,
\label{eqn:deltaDMbroke}
\end{equation}
where the {\it{total}} \ DM$_{\rm PSR,\ j}$  and RM$_{\rm PSR,\ j}$ 
are observed and each term in the summations is modeled. We name the procedure embodied by the above two equations as the ``arm-by-arm'' technique.

As Eqs.  \ref{eqn:deltaRMbroke} and \ref{eqn:deltaDMbroke}
 provide the means to determine  $\delta \rm{RM}_{\rm j,k}$ and $\delta \rm{DM}_{\rm j,k}$, Eq.  (\ref{eqn:BcosTheta}) can be rearranged 
to yield the magnetic field magnitude $|B_{\rm k}|$ in the farthest ($k^{th}$) subpath along this particular ($j^{th}$ 
pulsar's) LOS:

\begin{equation}
|B_{\rm j,k}|  = 1.232\ \frac{  \delta \rm{RM}_{\rm j,k} /  \cos \theta_{\scriptsize{\texttoptiebar{(B\ ds)}_{\rm j,k} } } } { \delta \rm{DM}_{\rm j,k} }\  \mu\rm{G}.
\label{eqn:Bmag}
\end{equation}
The numerator of the above equation, {$(\delta \rm{RM}_{\rm j,k} /  \cos 
\theta_{\scriptsize{\texttoptiebar{(B\ ds)}}_{\rm j,k}}$}) 
will hereafter be called the ``geometrically-corrected $\delta \rm{RM}_{\rm j,k}$.''

To our knowledge, our work represents the first effort to geometrically correct all analyzed RMs, 
thereby enabling systematic magnetic analyses along large segments of spiral arm and interarm regions, even 
while $\cos \theta_{\scriptsize{\texttoptiebar{(B\ ds)}}}$ varies significantly.

In much of what follows, we deploy Eqs. (\ref{eqn:deltaRMbroke}) through  (\ref{eqn:Bmag}) as our fundamental 
 analysis tools on a given pulsar LOS in a particular zone. A collective version of Eq. (\ref{eqn:Bmag})
 that averages the contributions of all pulsar - Earth  
subpaths originating at pulsars in the $k^{\rm th}$ zone and ending at the Earthward  edge of that zone, 
provides a mean value of the arm-aligned (or anti-aligned) magnetic field in that zone:
 
 \begin{equation}
\overline{ |B_{\rm k}| } = 
1.232\ \overline{ \left( \frac{  \delta \rm{RM}_k /  
\cos \theta_{\scriptsize{\texttoptiebar{(B\ ds)}} _k} } 
{\delta \rm{DM}_k }\right) } \  \mu\rm{G};
\tag{10a}
\label{eqn:Bmagmean}
\end{equation}
where the overline here represents an averaging process over all  $j=1,2,3,...,N_k$  LOS 
subpaths originating in zone $k$; i.e, those subpaths originating from all pulsars lying in zone
$k$. (The specific nature of the ``averaging process'' will
be delineated below.)

However, the helicity of $\cos \theta_{\scriptsize{\texttoptiebar{(B\ ds)}} _k}$ is {\it{a\ priori}} unknown. Thus, Eq. \ref{eqn:Bmagmean} does not tell us whether the field in a given magnetic arm is CW or CCW. 
Let us define a CCW  $k^{\rm th}$ arm tangential unit vector $\hat{t}_k$. Then, ${\bf{B}}_{\rm k}=h_{\rm k}   |B_{\rm k}| \hat{t}_{\rm k}$ with a CCW ($h_k=+1$) or CW ($h_k=-1$) helicity. We can then replace
$\cos \theta_{\scriptsize{\texttoptiebar{(B\ ds)}} _k} $ with 
$ h_k \cos \theta_{\scriptsize{\texttoptiebar{ (t\ ds)}} _k}  $ such that we can solve for both the magnitude ($\overline{ |B_{\rm k}| }$) and helicity ($h_k$) of the magnetic field in a given zone e.g.,

 \begin{equation}
h_k \ \overline{ |B_{\rm k}| }= 
1.232\  \overline{ \left(  \frac{  \delta \rm{RM}_k /  \cos \theta_{\scriptsize{\texttoptiebar{(t\ ds)}} _k} } {\delta \rm{DM}_k } \right) }  \  \mu\rm{G}.
\tag{10b}
\label{eqn:Bmagmeanandhelicity}
\end{equation}

Under our working assumption that the  magnetic field magnitude $|B|_{\rm j,k}$ is approximately 
constant in zone $k$ for all $j=1,2,3,...N_k$ pulsar  LOS originating in the zone,  Eq. \ref{eqn:Bmag} indicates that there will be
a linear relationship between the geometrically-corrected $\delta \rm{RM}_{\rm j,k}$ and  $\delta \rm{DM}_{\rm j,k}$. We therefore create $N_k$ ordered pairs, $z_{j,k}$,  of such denominators and numerators for zone $k$, where: 
\setcounter{equation}{10} 
\begin{equation}
 z_{j,k}=(\delta \rm{DM}_{\rm j,k},\ \ \delta \rm{RM}_{\rm j,k} /  \cos \theta_{\scriptsize{\texttoptiebar{(B\ ds)}}_{\rm j,k}}),
 \label{eqn:orderedpair}
  \end{equation}
  
\noindent with $j=1,2,3,...,N_k$. Our adopted ``averaging process'' for determining  $\overline{ |B_{\rm k}| }$, the mean magnetic field magnitude in the $k^{th}$ zone, takes advantage of this linear relationship by fitting a best slope to the set of $N_k$ pulsars' $z_{j,k}$  within the $k^{th}$ region and then substituting that best slope for the overlined quantities in Eqs. 10. In the fit, we do {\it{not}} weight each $z_{j,k}$  by its  observational uncertainties, since spatial fluctuations in the interstellar medium 
itself induce RM variations of unknown  amplitude that tend to be significantly 
larger than instrumentally-induced 
RM uncertainties \citep{Ohno1993, Sunet2008}.  Instead, we weight each $z_{i,k}$ equally in the fit after 
rejecting  $N_{k,r}$ LOS which represent extreme versions of the local 
variation phenomenon, nonphysical (negative) $\delta$DMs\footnote{The nonphysical, negative $\delta$DMs is solely due to the selected gridding for the spiral arms.}, or LOS almost perpendicular to the arm. 

We graphically depict this zone-averaging process for each region in a so-called ``ramp-plot'' where, for the selected ($k^{th}$) zone, we display the $N_k$  values of $z_{i,k}$ within that zone along with the line that best fits all of them.  Furthermore, as indicated by Eq.  (\ref{eqn:Bmagmeanandhelicity}), the {\it{direction}} of the  field is  also revealed by  the ramp-plot: a CCW  field along the arm would result in the best-fitting line having a positive slope, whereas a  CW field would lead to a negative slope.  

We apply a bootstrap process \citep{Efron1991} to estimate a first-order uncertainty in the fitted slope and in the resulting mean $B$-field\footnote{For the remainder of this work, all referred to B-fields have been geometrically corrected according to Eq. \ref{eqn:Bmagmeanandhelicity}.}  for the arm. For each such zone except the Local Arm, an additional uncertainty due to the covariance of the resulting slope with other arms' slopes was calculated and included in our error budget.\footnote{The Local Arm is excluded from this analysis since there are no nearer arms to influence its LOS.} The additional term was negligible in all cases except that of the Perseus Arm, but is nevertheless included in all $B$-field uncertainties reported in this paper.

Note that while Eqs.(\ref{eqn:deltaRMbroke}) - (\ref{eqn:orderedpair}), as displayed, will reveal properties of the farthest ($k^{th}$) zone from Earth, similar versions can be deployed to study nearer zones; indeed the farther zones can only be studied {\it{after}} the nearer zones have been evaluated and modeled, as the near-zone  RM and DM contributions  must be subtracted from the observed values  to yield the farthest zones' portions ({\it{cf.}} Eqs. \ref{eqn:deltaRMbroke} and \ref{eqn:deltaDMbroke}, respectively.)  We use the YMW17 electron density model to calculate the 
 $\delta$DM$_i$ for any $i^{th}$ subpath along the LOS via a subpath piece of Eq. (\ref{eqn:DMdef}). However, there does not yet exist a reliable global magnetic field model that would enable us to perform a similar $\delta$RM$_i$ calculation with Eq. (\ref{eqn:RMdef}). Instead, we 
 develop our own model of the systematic magnetic field in a $k^{th}$ zone by 
solving for the intervening zone's {\bf{${\bf{B}}_i$}} and hence associated $\delta \rm{RM}_i =\int_{i} n_e {\bf{B}}_i \cdot d{\bf{s}} / 1.232$. We then apply Eq. (\ref{eqn:deltaRMbroke}) to find the portion of the observed RM attributable to the farthest $(k^{th})$ subpath from Earth.

This zone-by-zone procedure allows us to  optimize our fit  for  the magnetic zone's width and field  strength in each arm or interarm region independently.  (See Table \ref{table:Bs} for a list of these quantities.) To our knowledge, this procedure has not been used before. In principle, it can be extended in the future to additional zones as new RM measurements become available.

While our assumptions of a constant field within a zone  pointing parallel to its outer boundary are inevitably oversimplified, we will show that existing pulsar RM and DM measurements are consistent with these assumptions in most cases. Specifically, our  magnetic field measurements along  {\it{individual}} LOS through a specific arm or interarm region,  calculated via Eq.  (\ref{eqn:Bmag}), are generally consonant with our measured {\it{average}} fields therein, as calculated from Eqns.  10 and the fits to $N_k$ sets of $z_{\rm j,k}$. However, we should also note that the \textit{random} fields tend to be comparable or even larger in magnitude than the systematic ones studied here that could contribute to our measured RMs \citep{Ohno1993, Sunet2008}. This could lead to significant jitter in the RMs for a given arm.

\section{Systematic Magnetic Fields in the Galactic Plane}
\label{sssec:systematic}

The next sections apply our new technique to various spiral arm zones. We first investigate the Local Arm, where we illustrate our analysis techniques in some detail. We then move  
farther away from Earth; first to zones successively closer to the Galactic center (the Carina-Sagittarius Arm,  
the Sagittarius-to-Scutum Interarm Region, and the Crux-Scutum Arm); 
and then outward to the Perseus Arm. While we also have some new RMs in the Norma-Outer Arm region, they are not sufficiently densely  placed for us to detect 
systematic magnetic field trends there. Therefore, delineation of the Norma-Outer Arm will be deferred to future analysis. 

\subsection{Local Arm} \label{sssec:local}

While the locations and pitch angles of most spiral arms are fairly well agreed-upon, there is still some
disagreement on these parameters for the Local Arm. Consequently, much  recent work on Galactic structure has focused on this arm. 
The YMW17 Local Arm Center, shown as a dashed red curve in many of figures in our work, 
uses the parameters of the \citet{HouHan2014} third ``4 + Local'' Arm model 
which was fitted to HII regions. Their Local Arm
presents as an angling bridge between the Carina-Sagittarius and Perseus Arms. 

The resulting  Local arm has a much lower putative pitch angle than that of  the other  spiral arms. However, a growing body of  VLBI and Gaia parallaxes \citep[e.g.,][]{Xuet2013, Xuet2018, Reidet2016, Reidet2019}, and a statistical study of various spiral tracers \citep{Grivet2017} argue for a Local Arm pitch angle similar to that of most other arms, such that the Local Arm is approximately {\it{parallel }} to its adjacent Carina-Sagittarius and Perseus Arms.

We therefore define the Local Arm  as the region lying between the outer boundary of the Carina-Sagittarius Arm and the inner boundary of the Perseus Arm, still extending azimuthally over the approximate range of the red-dashed YMW17 Local Arm and capped at both ends by lines of constant Galactocentric azimuth.  It is replaced off its low-longitude end by a Perseus-to-Sagittarius Interarm Region (see \S\ref{sssec:PersToSag}). We model the Local Arm's magnetic field to point parallel to its newly defined outer boundary, rather than to the YMW17 Arm center. For all these reasons, the YMW17 Local Arm {\it{center}} is shown as a {\it{dashed}} curve wherever it appears in our figures.

\begin{figure}[t]
\includegraphics[trim={1.1in   5.1in 0.7in  2.3in},clip, scale=0.52 ]{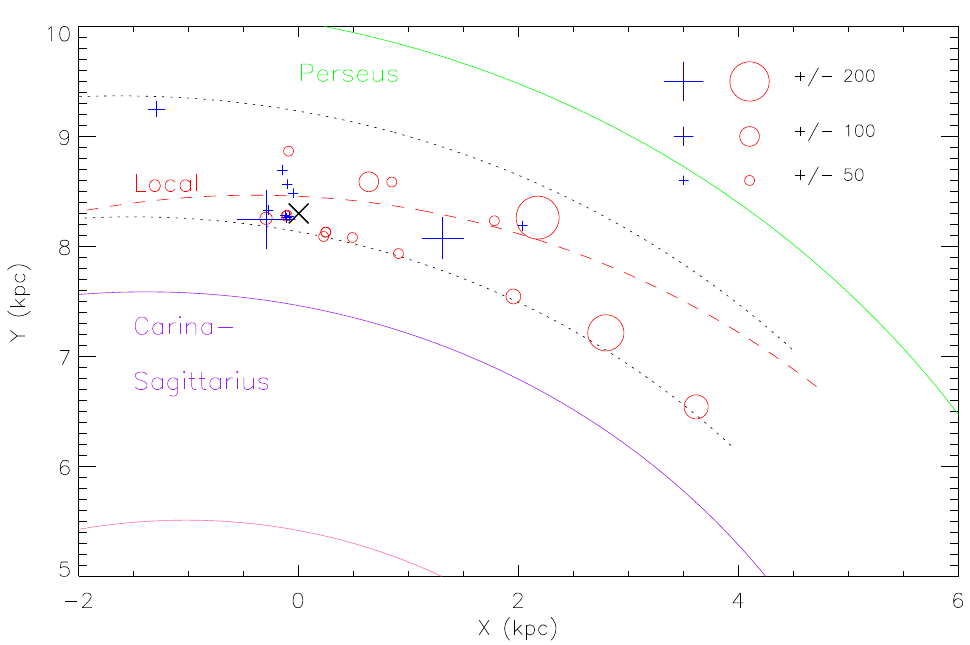}
\caption{ 
Measured RM$_{\rm PSR}$ of each low-latitude pulsar in our sample in the Local Arm region, except those in and beyond the Gum Nebula.$^9$  The black dotted curves delineate the nominal inner and outer boundaries of the (magnetic) Local Arm derived from our analyses. A  Perseus-to-Sagittarius  Interarm Region (not shown) is presumed to extend below the low-longitude end of the Local Arm (see \S\ref{sssec:PersToSag}). We model the magnetic field within the Local Arm to be parallel to our  Local Arm's outer boundary. The black $\times$ symbol at $(x,y) = (0.0, 8.3)$ marks the location of the Solar System.}
\label{fig:Local}
\end{figure}

We adjusted the Local Arm’s Galactocentric inner and outer edges until they encompassed a zone of roughly constant 
magnetic field directed along the Arm, while simultaneously excluding zones of significantly different magnetic field. See Table \ref{table:Bs} for the parameters of our adopted magnetic inner and outer Galactocentric boundaries
for the Local Arm. In Fig.  \ref{fig:Local}, we illustrate the Galactic plane locations and measured RMs of all pulsars having measured values in the above-defined Local Arm region, except for those pulsars that lie within or beyond the Gum 
Nebula.\footnote{For these purposes, we adopt the YMW17 model of the location and size of the Gum  Nebula --- an ellipsoidal shell with its long axis oriented  perpendicular to the Galactic plane, centered   at a distance of 450 pc   from Earth, toward $(l,b)=( 264\degree,-4\degree),$ with a radius  of 126.4 pc in the plane parallel to the Galactic plane that contains the center of the Nebula.} 

The Gum Nebula serves as a major localized perturbation to the DMs and RMs of pulsars located in or beyond it. While \citet{Purcellet2015} have created a model of the magnetic 
field in and near the Nebula 
itself, its uncertainties  suggest that it is best to remove pulsars in and beyond this region from our 
study of the Galactic magnetic field, rather than attempting to use a poorly constrained magnetic model of the Gum Nebula to calculate its contributions to larger-scale fields. The  pulsars  removed for this reason from Fig. \ref{fig:Local} and from our magnetic field fits are 
listed in Table A1. 

\begin{figure*}[t]
{
	\includegraphics[trim=1in   5.3in 0in  1in, scale=0.5, width=0.52\linewidth ]{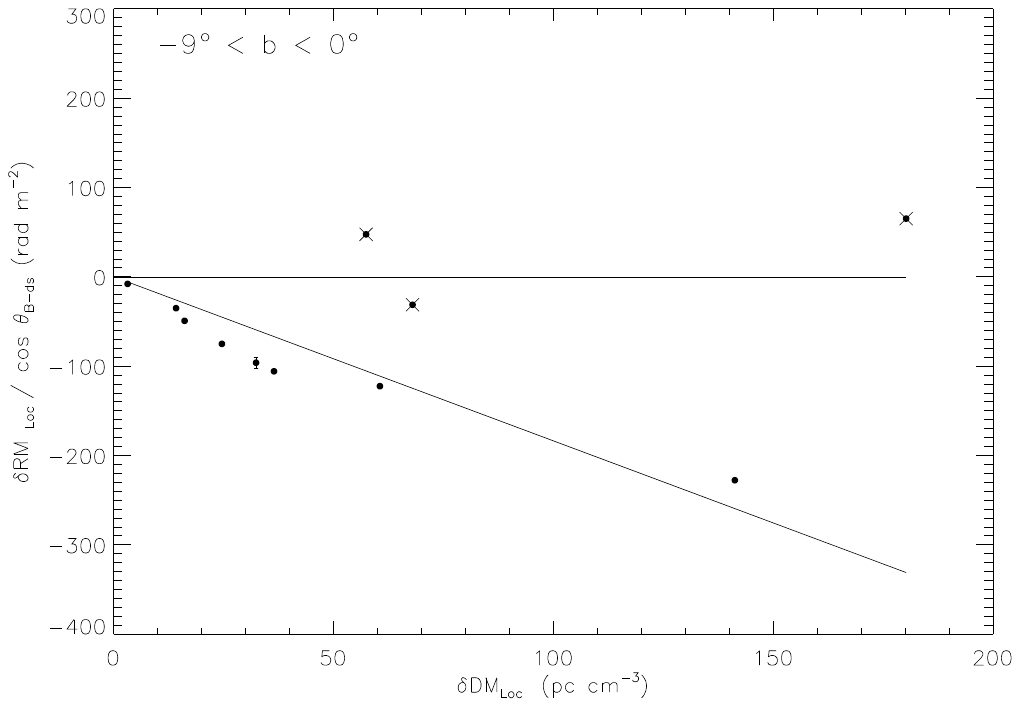} 
	\includegraphics[trim=1in   5.3in 0in  1in, scale=0.5, width=0.52\linewidth   ]{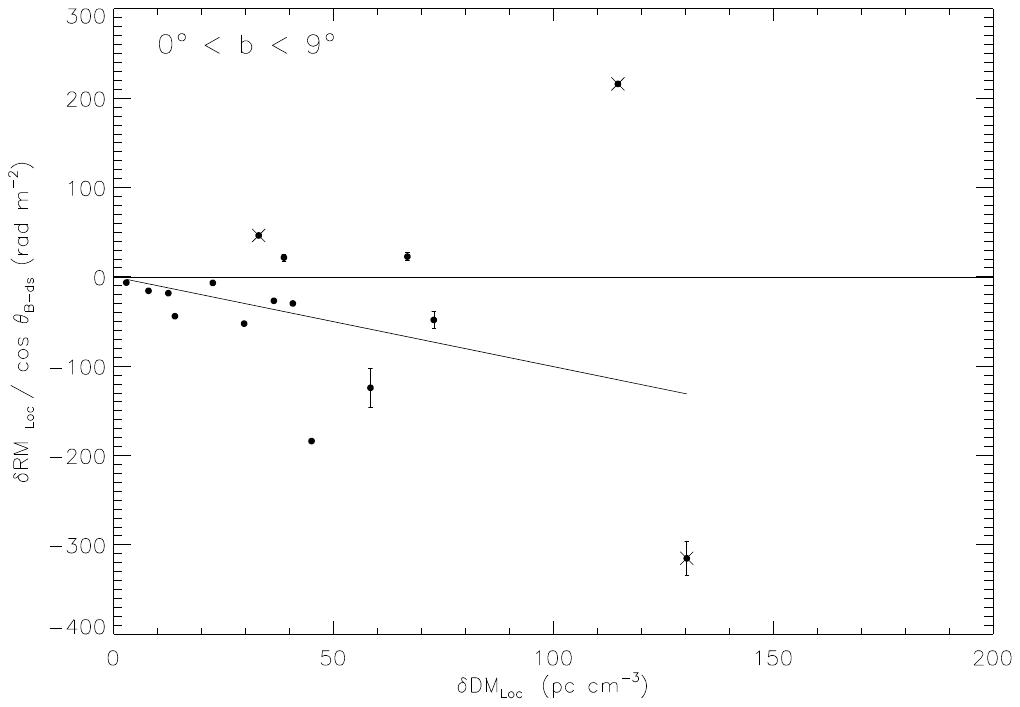} 
}
\caption{\textit{Left figure:}
Geometrically-corrected $\delta$RM$_{\textrm{Local}}$ as a function of $\delta$DM$_{\textrm{Local}}$ for  the Local arm pulsars shown in Fig. \ref{fig:Local} with negative Galactic latitudes. \textit{Right figure:} Same as the left except this plot is for pulsars with  positive Galactic latitudes. In each case, we show the best linear fit to the respective values. Outliers are shown with $\times$’s, with details listed in Table  \ref{table:outliers}. 
}
\label{fig:LocalRamp}
\end{figure*}

It is important to note that the Gum Nebula lies in a complicated region of the sky, which has led other investigators  \citep[e.g., ][]{XuHan2019} to create rather different models of its distance, size, and even shape from that of  \citet{Purcellet2015} and YMW17. Any of these alternate models, if adopted, would in turn  lead to a somewhat different ``Gum shadow zone of exclusion'' and resulting set of rejected pulsars. 

In Fig. \ref{fig:LocalRamp}, we plot the geometrically-corrected  $\delta$RM$_{\textrm{Local}}$ as a function of $ \delta \rm DM_{ \textrm{Local} }$ for all pulsars in the Local Arm, along with the resulting unweighted linear fit. We exclude from our fit pulsar LOS with extreme values of $\delta$RM$_{\rm Loc}$  / $\cos \theta$, $\delta$DM$_{\rm Loc}$, or $B_{\rm Loc}$. Table \ref{table:thresholds} presents the  rejection thresholds for each of these quantities in this Arm and all successively analyzed zones while Table  \ref{table:outliers} lists all ``outlier'' LOS thereby rejected from each zone's fit and the rationale for so doing.

We find a 2-3\hspace{0.2em}$\sigma$ difference in the respective linear fits for pulsars below and above the Galactic plane, and hence Fig. \ref{fig:LocalRamp} is split into two sub-plots: one for negative Galactic latitude pulsars and another for those at positive Galactic latitudes. For negative (positive) Galactic latitudes, the slope of the best-fit line is $-2.0 \pm 0.3 $ ($-1.0 \pm 0.5$), which implies a magnetic field  of magnitude $2.5 \pm 0.4$ ($1.3 \pm 0.6$) $\mu$G, pointing CW and parallel to the Arm's outer boundary. 

We also study the Local Arm's magnetic field along {\it{individual}} pulsar LOS as a function of both longitude and latitude. Other than a difference  above and below the Galactic plane consonant with the associated RM difference noted above, we do not find any other features that depend on longitude or latitude in this region.

To our knowledge, this work is the first  that indicates  a distinction between the Local Arm disk's magnetic field strength at  positive and  negative Galactic latitudes, albeit only at the 2-3\hspace{0.2em}$\sigma$ level. A similar north-south asymmetry has been noted in the Sagittarius Arm \citep{Ordoget2017, Maet2020}, and indeed we find such asymmetries in most zones that we analyze in this work. See \S\ref{sec:UpDown} for further discussion of the north-south asymmetries.

A CW magnetic field within the Local Arm has been well documented within the literature, with most recent investigators also suggesting that it points parallel to the nearby major arms, rather than having the anomalous pitch angle of the YMW17 model's Local Arm.  There are also several estimates of the field magnitude, starting with \citet{Manchester1972} who used pulsar RMs to infer a local field of $\sim3.5~\mu$G , directed toward $\ell\sim90$\degree. \citet{Sunet2008}, using a wide variety of radioastronomical measurements but no pulsar RMs, modelled the Local Arm parallel to the nearby major arms and found that the magnetic field magnitude  is $\sim2\mu$G. 

Most recently, \citet{Xuet2022} 
fitted straight lines to sets of pulsar's total RMs as a function of pulsar distance in low-latitude, $\sim6\degree$ longitude wedges in the first Galactic Quadrant, deriving LOS field strengths where possible. The wedges most comparable to ours are that of $62\degree < l < 70$\degree  and $80\degree < l < 90$\degree, as they include pulsars in the Local Arm. They find LOS magnetic field magnitudes of 1.9 $\pm$ 0.7 $\mu$G and 2.8 $\pm$ 0.6 $\mu$G, respectively, for the two ranges. As illustrated by  our Eq. \ref{eqn:Bmag}, these values should be geometrically corrected in order to yield field magnitudes along the Local Arm rather than along the LOS. This correction would boost the field strength by $\sim 10$\% for the lower-longitude wedge, and by a  negligible amount for the higher-longitude one. 
Note that unlike our arm-specific fits, their longitude-wedge fits only incorporate RMs over relatively narrow longitude ranges. The error bars on their two Local Arm fits  (after geometrical correction) plus ours all overlap.

\subsection{Local-to-Sagittarius and Local-To-Perseus Interarm Regions}
\label{sssec:abut}

Our modeling, combined with the somewhat limited number of pulsar RMs within the 
Local-to-Sagittarius and Local-To-Perseus Interarm Regions, indicates no current 
need for distinctive field configurations
(e.g., nulls,  additional  reversals, or significantly different magnitudes) in these zones. 
Consequently we modeled the Local Arm as directly abutting the Sagittarius and Perseus Arms.

\subsection{Sagittarius(-Carina) Arm } 
\label{sssec:Sag}

\begin{figure}[t]
\includegraphics[trim={1.1in   5in 0in  1.1in}, clip,scale=0.7 ]{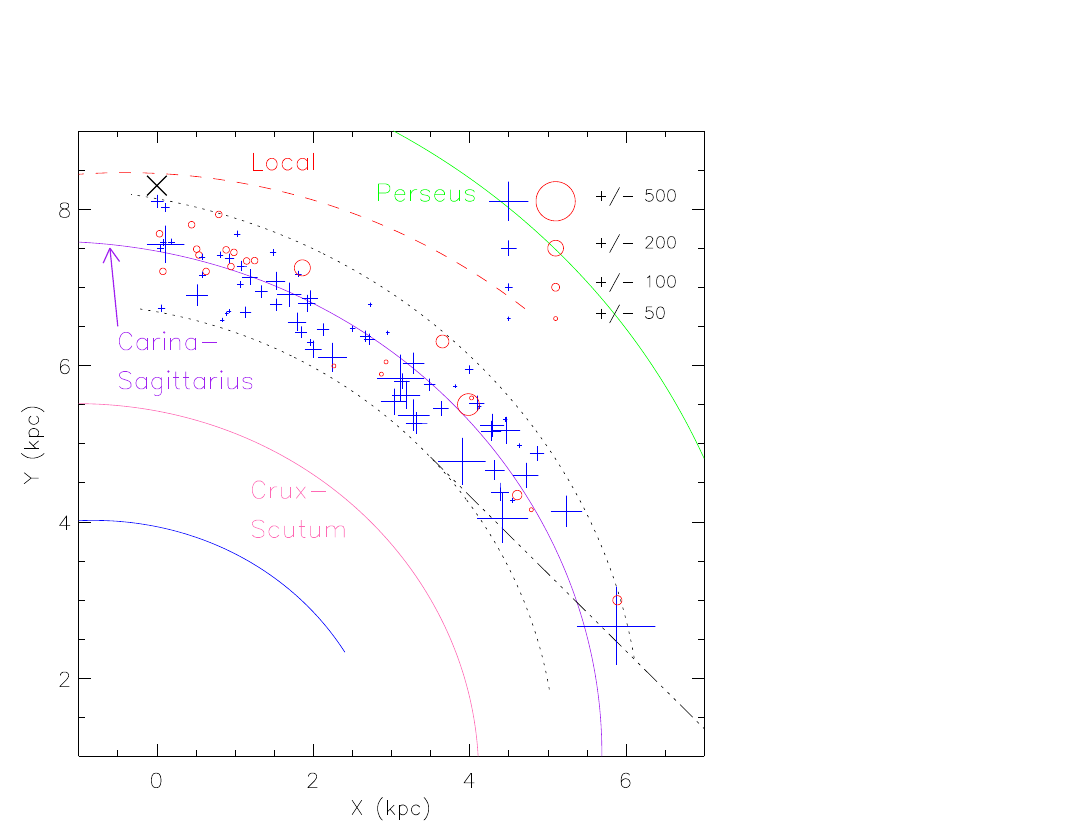}
\caption{Same as Fig. \ref{fig:Local}, except here for the Sagittarius Arm. The dot-dashed line represents the Galactic longitude limit below which the LOS would also sample the next inward region, the Sagittarius-to-Scutum Interarm.} 
\label{fig:Sag}
\end{figure}

Since the Arecibo telescope could access major portions of this arm in the first 
Galactic quadrant (defined as $0 \le \ell \le 90 \degree$)  
but not in the fourth, we study only the Carina-Sagittarius Arm pulsar LOS lying in 
the first quadrant, where the arm is called merely the 
Sagittarius Arm. Additionally, we eliminate the Sagittarius Arm pulsars lying below
the dot-dashed line in Fig. \ref{fig:Sag} (i.e., those whose LOS intersect the 
Sagittarius-to-Scutum Interarm Region along the path to Earth) since the determination of the Interarm magnetic properties 
itself depends on the magnetic properties of the Sagittarius Arm, thereby creating a situation 
in which the two zones' inferred magnetic properties would depend on each other. 

As explained above in \S\ref{sssec:abut},
the Galactocentric outer edge of the Sagittarius Arm is identical to the Galactocentric 
inner edge of the Local Arm. For the Sagittarius Galactocentric inner arm boundary, we 
followed the same procedure as that for the Local Arm, adjusting the boundary until the 
Sagittarius Arm encompassed a zone of roughly constant  magnetic field directed 
along the Arm. See Table \ref{table:Bs}  
for the parameters of our adopted  inner and outer Sagittarius  Arm boundaries.  All pulsars 
meeting the above criteria, and their measured total RMs, are shown in Fig.  \ref{fig:Sag}.

\begin{figure*}
{
	\includegraphics[trim=1.01in   5.33in 0in  1in, scale=0.4, width=0.52\linewidth ]{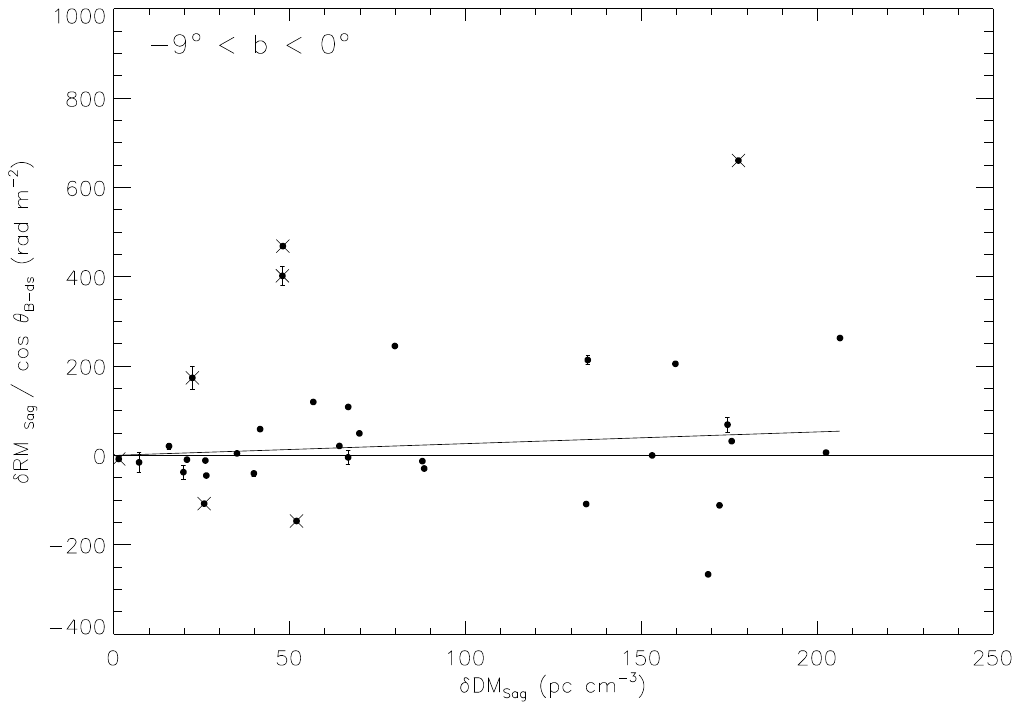}  
	\includegraphics[trim=1in   5.3in 0in  1in, scale=0.4, width=0.52\linewidth   ]{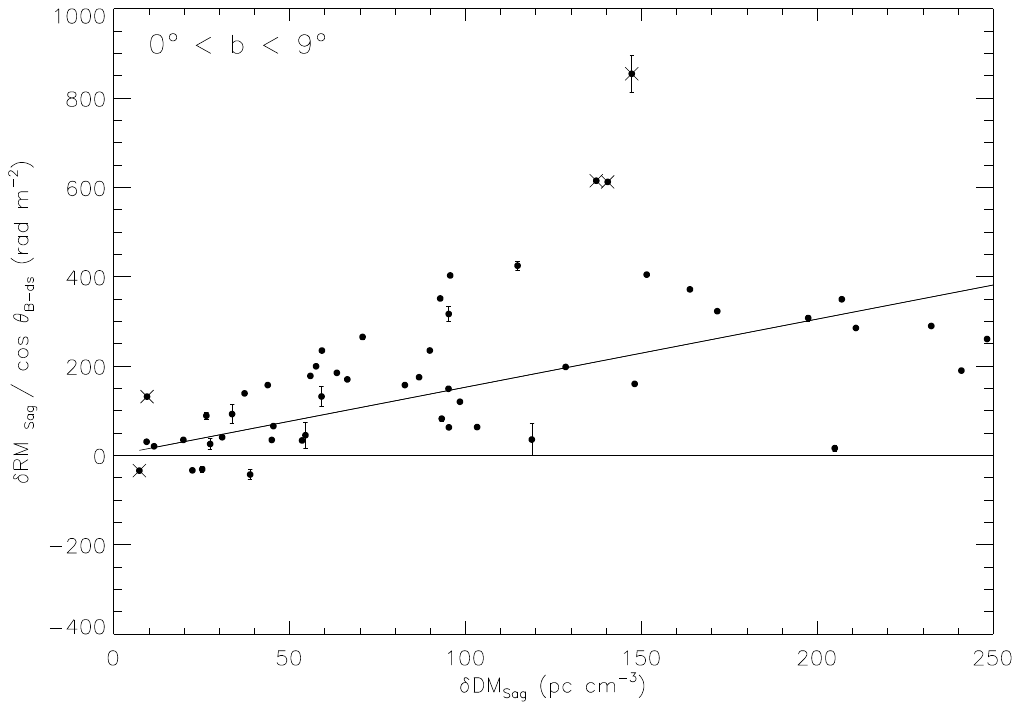} 
		 
}
\caption{Same as Fig. \ref{fig:LocalRamp} except for pulsars in the Sagittarius Arm. 
}
\label{fig:SagRamp}
\end{figure*}

The Sagittarius Arm components of geometrically-corrected $\delta$RM and of $\delta$DM  
for each of these pulsars were calculated from Eqs.   \ref{eqn:deltaRMbroke} and 
\ref{eqn:deltaDMbroke}, respectively, so as to remove the 
Local Arm contributions to the observed total quantities RM$_{\textrm{PSR}}$ 
and DM$_{\textrm{PSR}}$. Given that
we found different slopes and hence different magnetic fields in the intervening
positive- versus negative-latitude Local 
Arm LOS, we also split the Sagittarius Arm's 
geometrically-corrected $\delta$RMs and $\delta$DMs into positive- and negative-Galactic latitude sets. In Fig. \ref{fig:SagRamp}, we show the geometrically-corrected $\delta \rm{RM}_{\textrm{Sag}}$ as a function of ${\delta\rm DM}_{\rm Sag}$, and the resulting unweighted linear fit to these two datasets. Under our working assumption that the magnetic field magnitude in the Sagittarius Arm is constant and directed parallel or antiparallel to it, we use Eq. (\ref{eqn:Bmagmeanandhelicity}) to solve for the  magnitude and direction of ${\rm B}_{\rm Sag}$ along the specified fitted region of the Arm. 

The resulting best-fit slopes for negative (positive) Galactic latitude pulsars in the Sagittarius Arm is $0.3 \pm 0.3$ ($1.5 \pm 0.2$), which implies a magnetic field strength of $0.3 \pm 0.3$ ($1.9 \pm 0.2$) $\mu$G, both pointing in the CCW direction along the Arm.  We note that our negative Galactic latitude fit, while significantly different from the positive-latitude value, is not statistically different from zero. 
There are several plausible explanations --- e.g., the {\it{random}} or locally perturbed magnetic field might dominate here  over its systematic component  \citep{Ohno1993, Sunet2008}; our model might not properly describe the overall magnetic field in this region (see, e.g., our evidence in \S\ref{sssec: field reversal sag arm} for a field reversal between the inner and outer galactocentric boundaries of  the Arm itself); or the number of measurements is simply inadequate to achieve statistical significance among the fluctuations (which implies that additional measurements might permit a more robust fit).

\begin{figure}[t]
\includegraphics[scale=0.50,trim={1.0in 5.0in 0.0in 1.7in},clip]  
{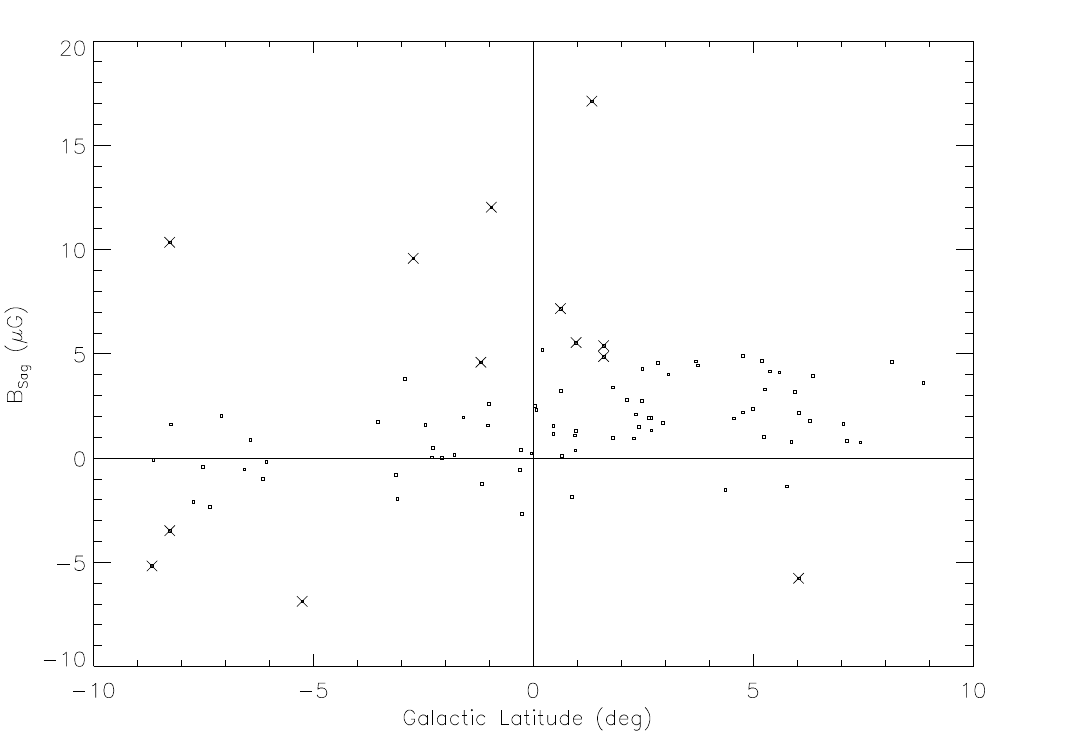}
\caption{Derived longitudinal magnetic field strengths (positive is CCW) along our Sagittarius Arm zone as a function of Galactic {\it{latitude}}. Outliers are denoted with an $\times$ and listed in Table \ref{table:outliers}. } 
\label{fig:SagBLatitude}
\end{figure}

The difference between  our $\delta$RMs above and below the plane in the Sagittarius Arm is also visible in Fig. \ref{fig:SagBLatitude}, where we show the Sagittarius Arm B-fields determined from individual pulsar LOS as a function of Galactic latitude. A north-south asymmetry (both in the Local and Sagittarius Arms) is echoed in previous Galactic synchrotron emission and total RM or Faraday Depth (FD)\footnote{In the simplest cases, RMs and FDs are identical.} measurements of pulsars and EGS in these directions (\citet{Ordoget2017}\footnote{These results will be further detailed in \S\ref{sssec: field reversal sag arm}.}, \citet{Maet2020}). The later of these references attributed differences in the RM and magnetic field above and below the plane in the longitude region of $ 35\degree < l < 50\degree$ to an odd z-parity field within the disk along the Sagittarius arm. We discuss this asymmetry in more detail in \S\ref{sssec: field reversal sag arm} in the context of the field reversal inside the Solar circle.

\cite{Han2018} and \citet{Xuet2022} also studied sections of the low-latitude Sagittarius Arm that we analyzed. Their field estimation technique was reminiscent of ours, in that they fitted a straight line to  RM as a function of DM for a set of pulsars. However, unlike our procedure, they used the {\it{total}} observed RM and DM rather than first extracting only the Sagittarius Arm component. They also did not geometrically correct the RMs. These issues were partly ameliorated by their confining each of their analyses to the set of pulsars lying within a relatively narrow-longitude wedge of the Galactic plane (e.g., the longitude wedges discussed in \S\ref{sssec:local}), rather than attempting a fit to all pulsars within a significantly longer region of an arm as we do. 
 
\citet{Han2018} found a magnetic field strength  of 1.4 $\pm$ 1.0 $\mu$G in the Sagittarius Arm over the longitude range of 45$\degree$ to 60$\degree$ between 3 and 8.5 kpc from Earth, where the LOS traverses the Sagittarius Arm over relatively long distances. Despite the above-noted limitations on their technique, their (relatively large) error bar overlaps those of our north and south fits.

None of the three  low-Galactic-latitude Sagittarius Arm longitude wedges studied by \citet{Xuet2022} correspond well to the long Sagittarius Arm segment considered in this work, so direct comparison of our calculated Sagittarius magnetic field with each of their fits is difficult. Nevertheless, we attempt to do so in what follows. 

(i) Their ``Inner Sagittarius'' fit covers  the region from  $44 \degree < l < 50 \degree$ and distances ranging between 3 and 12 kpc. LOS in this zone traverse the Sagittarius-to-Scutum Interarm Region for their first 2 kpc, and then obliquely cross 
the Sagittarius and Perseus  Arms, rendering those other non-Sagittarius Arm regions 
partly  responsible for the total RMs and DMs to which these authors fit.  
Also, as they do not geometrically correct their RMs along the LOS, their resulting  field values represent only the LOS component.
They find a field magnitude of 3.9 $\pm$ 1.3 $\mu$G, well above either our north or south fits for the field along the Arm. Additionally, as discussed below, there is a {\it{localized}} enhancement of the field in their sampling zone, which could also explain the discrepancy.
(ii) The authors' ``mid-Sagittarius'' fit almost entirely samples the Perseus Arm and does not include the  Sagittarius Arm. (iii) Their ``outer-Sagittarius'' fit samples  the region from  $56 \degree < l < 62 \degree$ 
and distances ranging from 3 to 8.5 kpc. Only the first two kpc of this wedge lie in the Sagittarius Arm. Their fitted field in the ``Outer Sagittarius'' region, $0.7 \pm 0.3\ \mu$G, is more in line with the mean of our north and south fits for this Arm, despite all of the differences between our fitting methods.

\citet{Shanahanet2019} (hereafter S19)  studied extragalactic 
RMs in the $\ell = (39 - 52)\arcdeg$ range near the Galactic plane, a region within 
our much broader (in Galactic longitude) Sagittarius Arm zone. Their techniques, unlike ours, did  not (and could not) explicitly split the LOS measurements they acquired into subpaths associated with particular Galactic zones (e.g., arms or interarm regions) due to the extragalactic nature of their studied sources. It is useful to compare and contrast S19's and our results. One of their most notable findings is a sharp and strong peak in measured extragalactic-source RM at  $ \ell  = (48 \pm 1)\arcdeg$, with the most extreme part extending over less than $1\arcdeg$ of longitude where RMs range up to more than 4000 rad m$^{-2}$. They argue  that this RM enhancement originates in material in or near the Sagittarius Arm. They also note that there is  {\it{no}} corresponding DM increase in {\it{pulsars}} at those longitudes,  which seems to imply (via Eq.  (\ref{eqn:Bmag})) that the arm's magnetic field is also greatly  enhanced in that region.  However, we note that Eq. (\ref{eqn:Bmag}) is not strictly applicable to S19's analysis for several reasons.  First, the  (Galactic) pulsars used by S19 are at various distances inside the  Galaxy so that the associated DMs represent path integrals of various shorter lengths, whereas their extragalactic-source RMs  are integrated over a much longer path. Additionally, the positions on the sky of  S19's disparate RM and DM sources are not identical. In an effort to further investigate this region, 
we carried out two related analyses.

\begin{figure*}
\includegraphics[trim=-0.1in 0.8in -0.2in  0.5in, scale=0.4, width=0.52\linewidth,clip]
{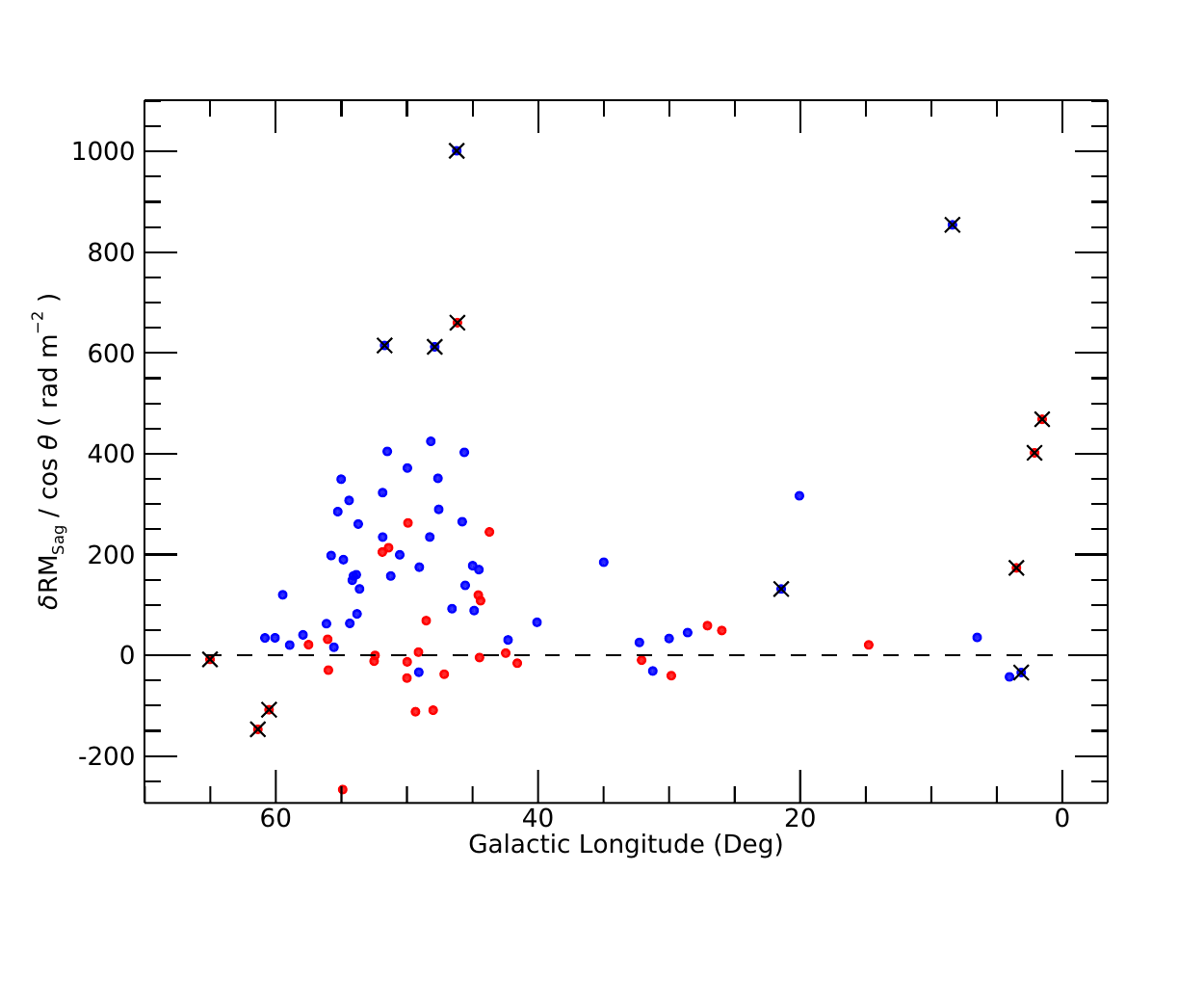} 
\includegraphics[trim=-0.1in 0.8in -0.2in  0.5in, scale=0.4, width=0.52\linewidth,clip]
{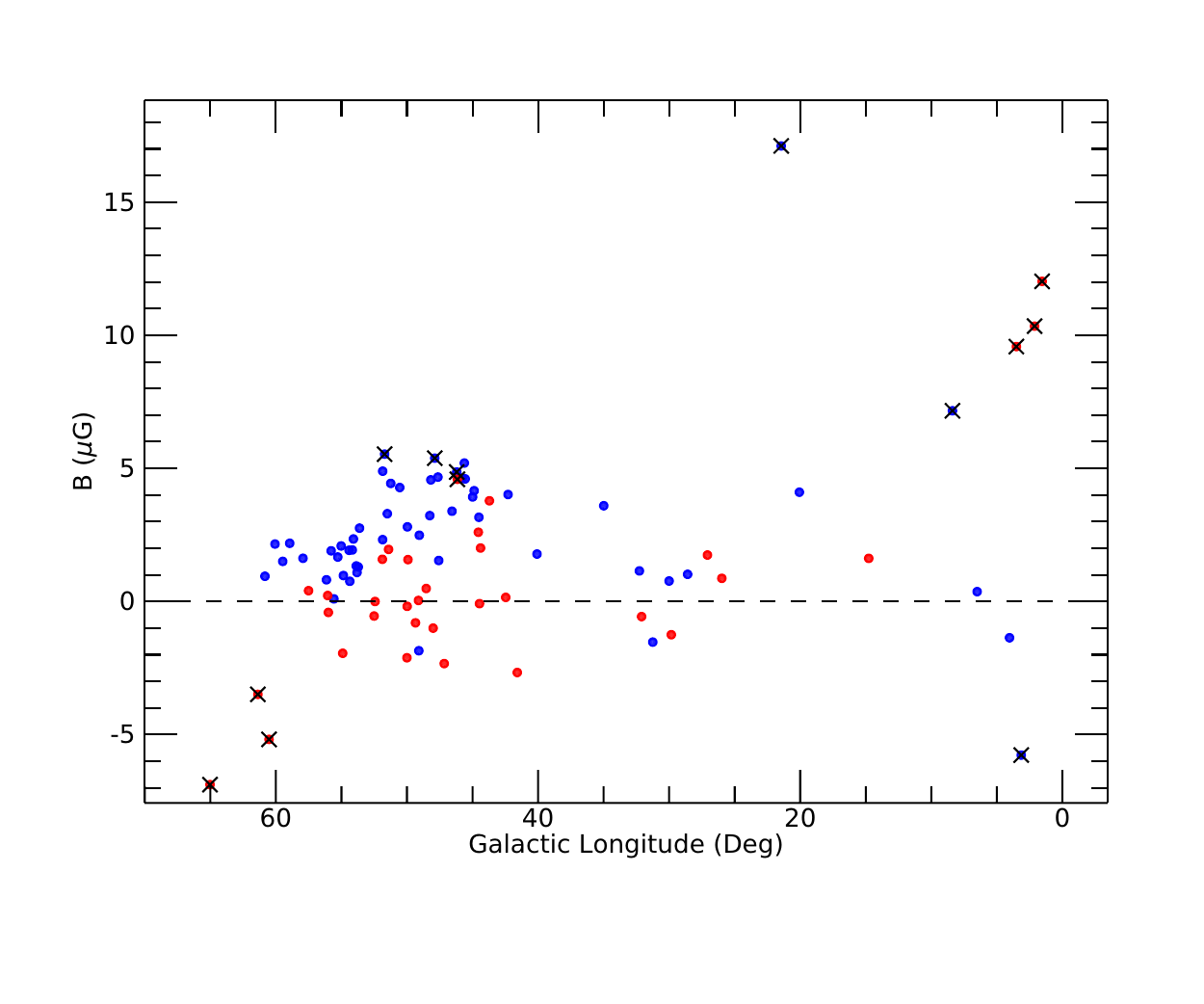} 
\caption{
\textit{Left figure:} The geometrically-corrected $\delta$RM$_{\rm Sag}$ as a function of Galactic longitude for the Sagittarius Arm  pulsars of Fig. \ref{fig:Sag}. \textit{Right figure: } Same as left panel, except that the magnetic field (positive is CCW) along the  Arm is displayed. Pulsar LOS at negative (non-negative) Galactic latitudes are given red (blue) symbols. Outliers are denoted with an $\times$ and listed in Table \ref{table:outliers}. }                    
\label{fig:SagRMBLongi}
\end{figure*}

First, we show our geometrically-corrected  $\delta$RM$_{\rm Sag}$ as a function of Galactic longitude for the Sagittarius Arm pulsars that are shown in Fig. \ref{fig:Sag} (see Fig. \ref{fig:SagRMBLongi}, \textit{left}). We {\it{do}} observe an enhanced geometrically-corrected $\delta$RM$_{\rm Sag}$ at the longitude of the S19 RM enhancement, although ours is broader in longitude and only about a quarter of their amplitude. Since the Arm field and the LOS are essentially parallel at this longitude, our geometrical correction makes little difference and cannot explain the discrepancy. We conclude that most of the significantly larger S19 extragalactic-source RM enhancement must actually occur in regions lying  {\it{beyond}} the Sagittarius Arm along the LOS, such as  the Perseus and/or  Outer Arms.  

Second, for each of the  pulsars that we used in our first study, we determine $B_{\rm Sag}$ \ via  Eq. (\ref{eqn:Bmag}) and then plot these values as a function of Galactic longitude (see Fig. \ref{fig:SagRMBLongi},  {\it{right}}.) While our measured $B_{\rm Sag}$ is generally enhanced by a factor of $\sim 2$ in this longitude range over our measured mean value along the much larger section of arm discussed  above ($0\arcdeg \lesssim \ell \lesssim 60\arcdeg$), this enhancement is much more modest than the claimed S19 RM increase. This puzzling discrepancy between the modest increase in our measured $B_{\rm Sag}$ and the large extragalactic-source RM enhancements of S19  has already been partially resolved above, when we showed that much of the latter enhancement must occur {\it{beyond}} the Sagittarius Arm. Additionally, because we can determine the geometrically-corrected  $\delta$RM$_{\rm Sag}$   
and also $\delta$DM$_{\rm Sag}$ along an identical LOS to a given pulsar, we can deploy Eq.  (\ref{eqn:Bmag}) much more accurately, and we find that much of the $\delta$RM$_{\rm Sag}$ enhancement in the numerator of Eq.  (\ref{eqn:Bmag}) is usually countered  by a  similar (though smaller) $\delta$DM$_{\rm Sag}$ enhancement in the denominator, thereby moderating any increase in the quotient, $B_{\rm Sag}$.

We also note that the enhancement in the geometrically-corrected  $\delta$RM$_{\rm Sag}$ around $ \ell  = (48 \pm 1)\arcdeg$ occurs principally at positive latitudes (Fig. \ref{fig:SagRMBLongi}, {\it{left}}). However, our sample of negative latitude pulsars within this region is limited, and hence a definitive conclusion cannot be made here.

It is also worth recalling that our fits for  $B_{\rm Sag}$  (see Fig. \ref{fig:SagRamp} and Table \ref{table:Bs}) assume and solve for a constant value along the Arm, whereas the modestly enhanced $B$ near $ \ell  = (50 \pm 5)\arcdeg$, especially at low positive latitudes, demonstrates that our model is an oversimplification in this region. It is beyond the scope of this work to use a more sophisticated model of the Arm's magnetic field.

 \subsection{Sagittarius(-Carina) to (Crux)-Scutum Interarm Region} 
\label{sssec:interSag2Scut}

\begin{figure}[b]
\includegraphics[trim={1.1in   5in 0in  1.2in}, clip,scale=0.79 ]{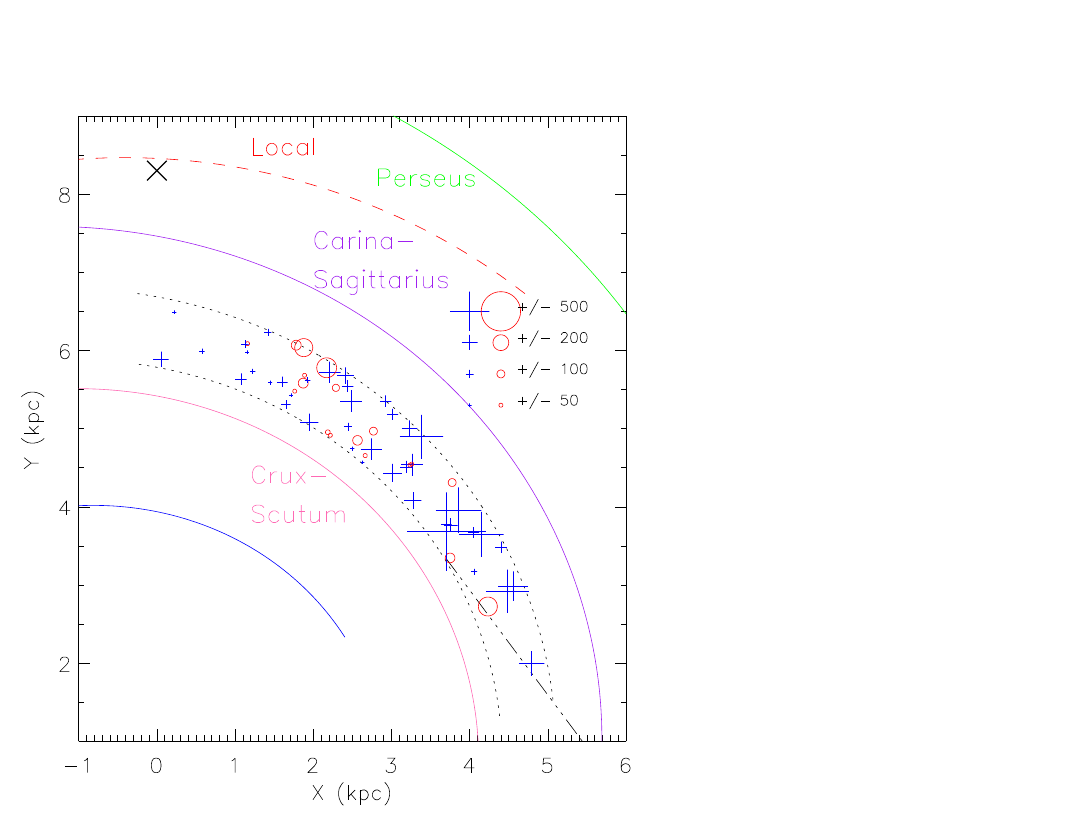}
\caption{
Same as Fig. \ref{fig:Sag}, except for the Sagittarius-to-Scutum Interarm Region. The dot-dashed line represents the Galactic longitude limit below which the LOS would also sample the next inward region, the Scutum Arm.} 
\label{fig:CarinaCrux}
\end{figure}

\begin{figure*}[t]
{
	\includegraphics[trim=1in   5.3in 0in  1in, scale=0.5, width=0.52\linewidth ]{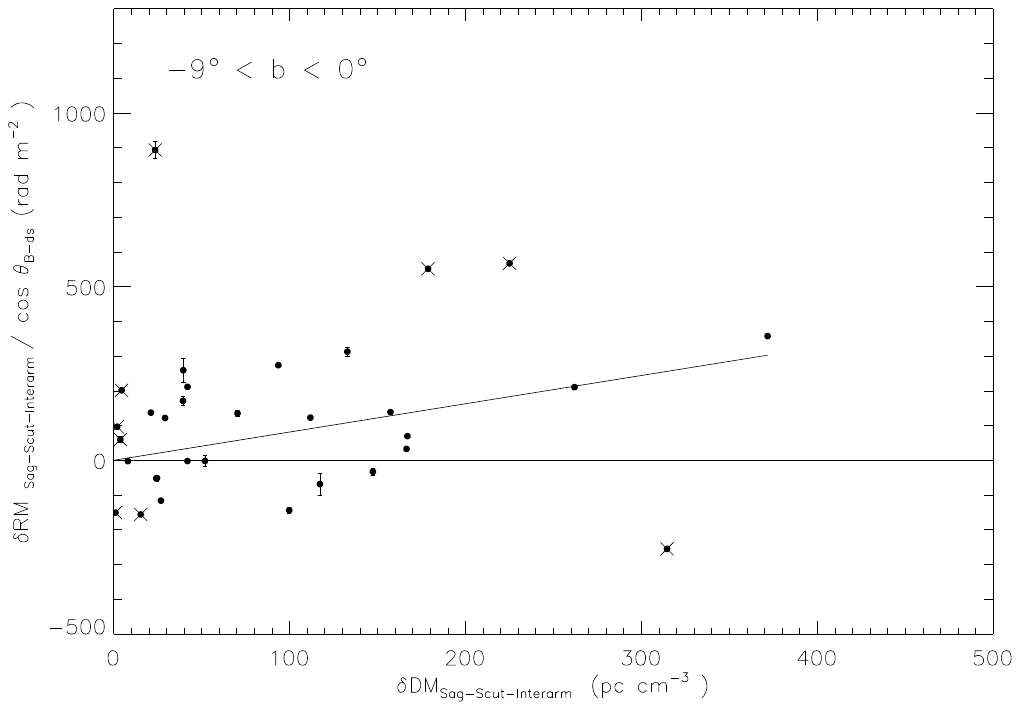} 
	\includegraphics[trim=1in   5.3in 0in  1in, scale=0.5, width=0.52\linewidth   ]{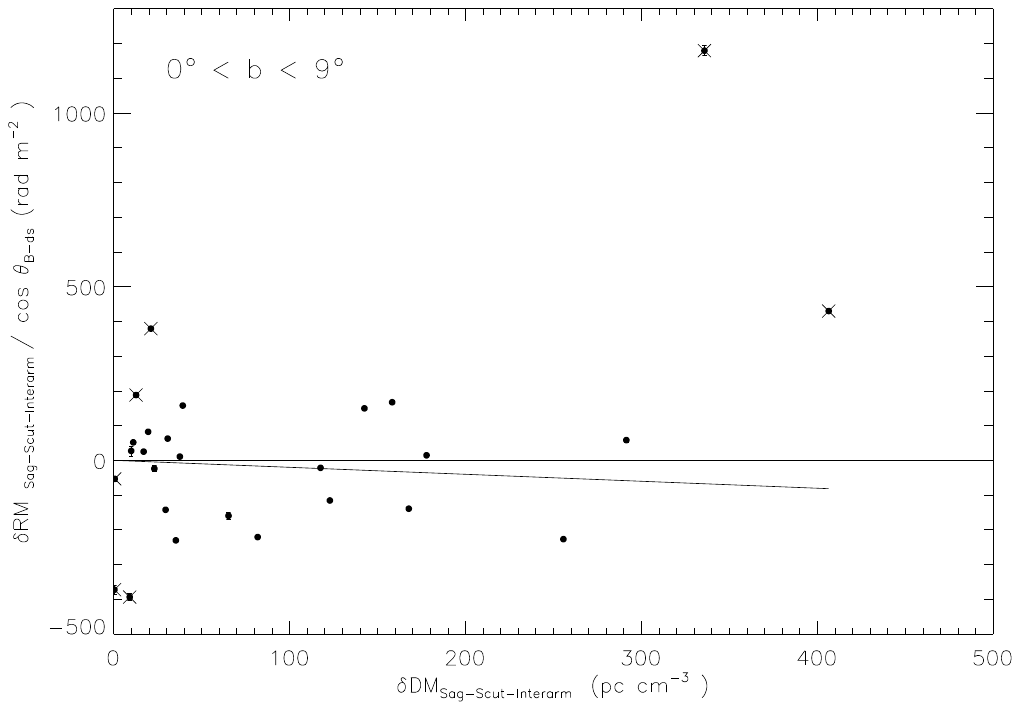} 
}
\caption{Same as Fig. \ref{fig:LocalRamp} except for pulsars in the Sagittarius-to-Scutum Interarm.}
\label{fig:SSIRamp}
\end{figure*}

The interarm region between the  Carina-Sagittarius Arm and the  Crux-Scutum Arm is also partly 
visible to the Arecibo telescope. We study
 it throughout  the first Galactic quadrant, where its name can be shortened to the 
 Sagittarius-to-Scutum Interarm Region, and will be further 
 shortened to ``SSI''  when used in a subscript. We define the {\it{outer}} Galactocentric boundary of this  interarm region to be identical to the inner boundary of the Sagittarius Arm.
To select the best inner boundary, we follow the same method that we used for the Sagittarius Arm in \S\ref{sssec:Sag}. Similarly to the Sagittarius Arm, we then reject pulsars lying between these inner and outer spiral boundaries whose pulsar-to-Earth LOS cross the  Scutum Arm so as to avoid a situation in which the two zones’ inferred magnetic properties would depend on each other. The line demarcating this final rejection criterion is shown in Fig. \ref{fig:CarinaCrux}, along with our chosen inner and outer spiral-shaped boundaries. Of the remaining pulsars, we determine $\delta$DM$_{\rm SSI}$ and the geometrically-corrected $\delta$RM$_{\rm SSI}$, plotting them separately in Fig. \ref{fig:SSIRamp} for  negative  and  positive Galactic latitude  pulsars.  

 As illustrated in Fig. \ref{fig:SSIRamp}, we find that the best-fit slope for negative (positive) Galactic latitude pulsars in the Sagittarius-to-Scutum Interarm Region is  $+0.8 \pm 0.2$ ($-0.2 \pm 0.3)$, which implies a magnetic field strength of $1.0 \pm 0.3$ ($-0.3 \pm 0.4$) $\mu$G in the CCW (CW) direction, parallel to the outer edge of the Sagittarius-to-Scutum Interarm.  While the negative-latitude fit is significant ($> 3\sigma$), we note  the significant scatter in  $\delta \rm{RM}_{\rm SSI}$ around $\delta \rm{DM}_{\rm SSI}$ = 0 (see the left subfigure of Fig. 
\ref{fig:SSIRamp}). This suggests that magnetic field variations near the Interarm's 
earthward edge are larger than usual.  For positive-latitude pulsars (see the right subfigure of Fig. \ref{fig:SSIRamp}), the best-fit slope and resulting systematic magnetic field are consistent with zero, presumably for one or more of the reasons we advanced to explain a similar result for the negative-latitude Sagittarius Arm pulsars (see \S\ref{sssec:Sag}). We also study the interarm's magnetic field along {\it{individual}} pulsar LOS, both as a function of longitude and latitude in this region. Similar to the Local arm, we do not find any features other than those noted above that depend on longitude or latitude in this region.

The CCW {\it{direction}} of the field that we derived  from negative-Galactic-latitude pulsars in this 
region agrees with the recent work of \cite{Han2018} and \citet{Xuet2022}. However, their measured 
magnetic field {\it{magnitudes}} of 3.3 $\pm$ 0.9 $\mu$G and 3.0 $\pm$ 0.9 $\mu$G, respectively, are significantly larger than our determinations at either negative or positive Galactic latitudes. 
As we  also noted in our Sagittarius Arm discussion (see \S\ref{sssec:Sag}), the earlier groups' methods differed from ours 
in that they did not geometrically correct their RMs, nor did they constrain their fits solely to the Arm in question. For example, the ``Scutum-Sgr'' fits of \cite{Han2018} and of \citet{Xuet2022} sample not only the interarm region, but also the zone extending 3-4 kpc beyond it, piercing both the Sagittarius and Perseus arms. An additional fit by the \citet{Xuet2022}, labeled ``Inner Sagittarius,'' also samples the interarm region but also obliquely intersects the Sagittarius and Perseus Arms, yielding the even higher magnetic field estimate of 3.9 $\pm$ 1.3 $\mu$G.  

\subsection{Scutum(-Crux) Arm} \label{sssec:Scutum}

\begin{figure}[t]
\includegraphics[trim={1.2in   5in 0in  1.15in}, clip,scale=0.8 ]{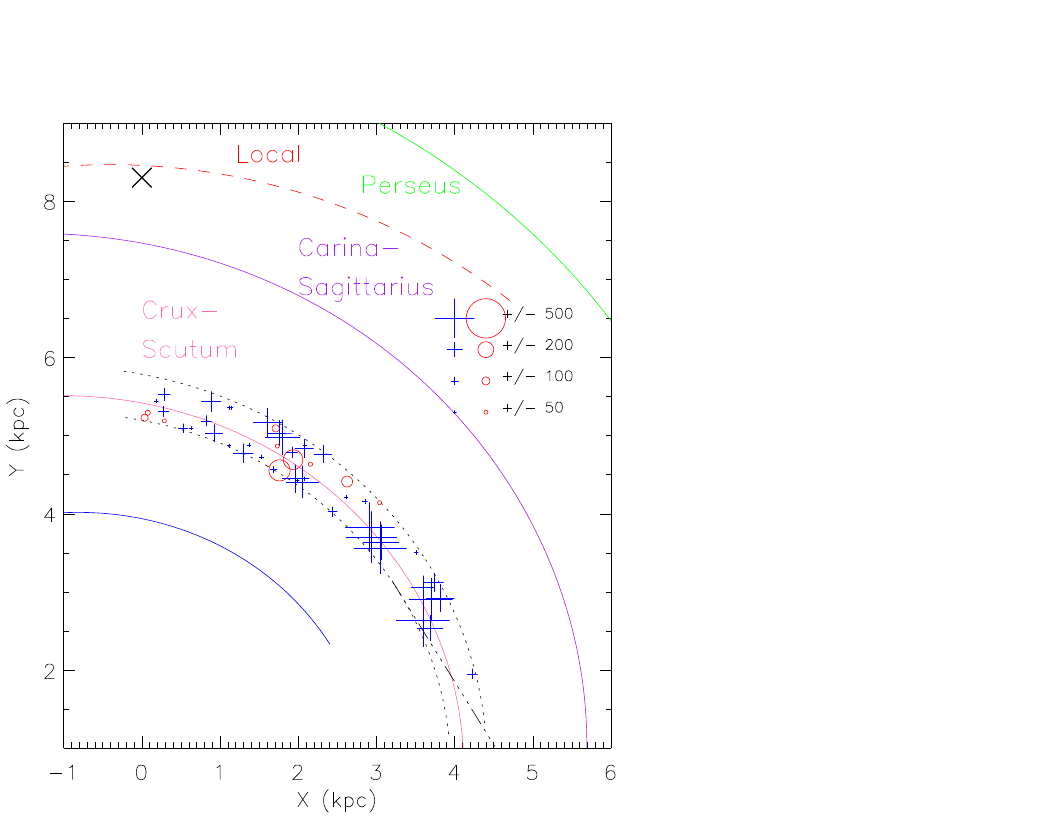}  

\caption{Same as Fig. \ref{fig:Local}, except for the Scutum Arm. The dot-dashed line represents the Galactic longitude limit below which the LOS would also sample the next inward region. }
\label{fig:Scut}
\end{figure}

\begin{figure*}[t]
\includegraphics[trim=0.7in   5in 0.4in  1in, scale=0.4, width=0.51\linewidth,clip ]{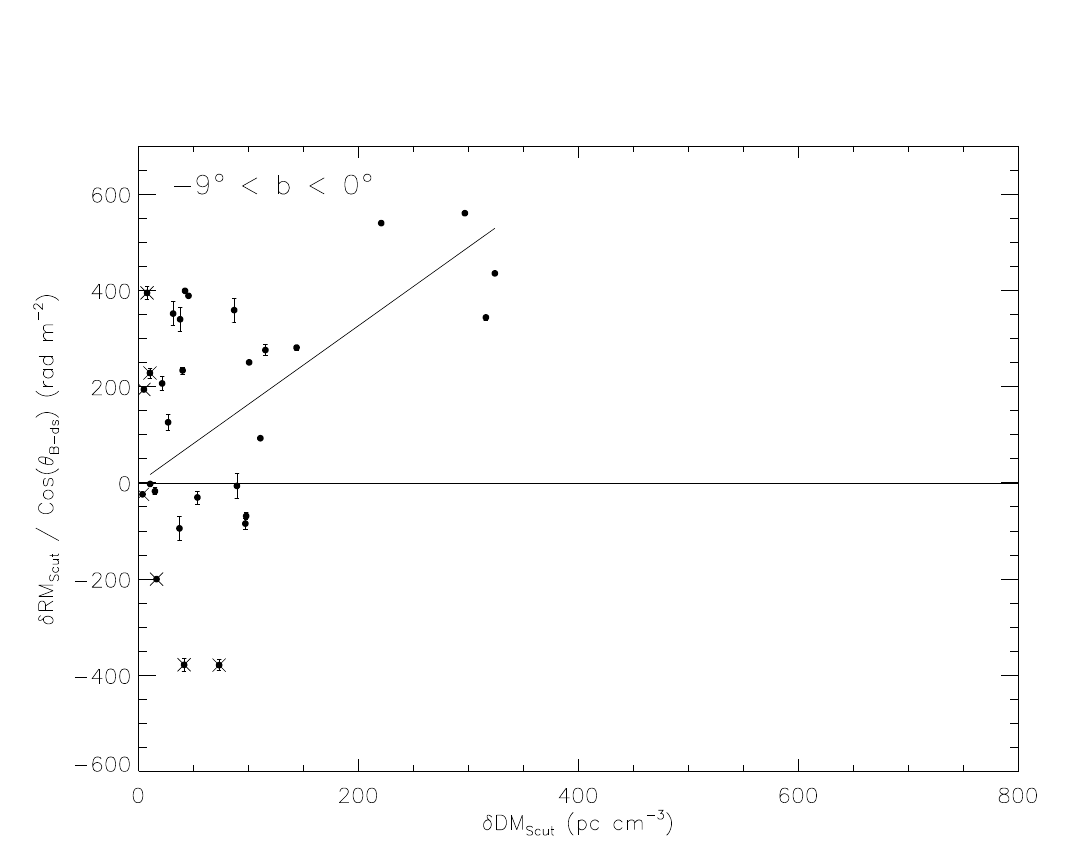} 
\includegraphics[trim=0.7in   5in 0.4in  1in, scale=0.4, width=0.51\linewidth,clip ]{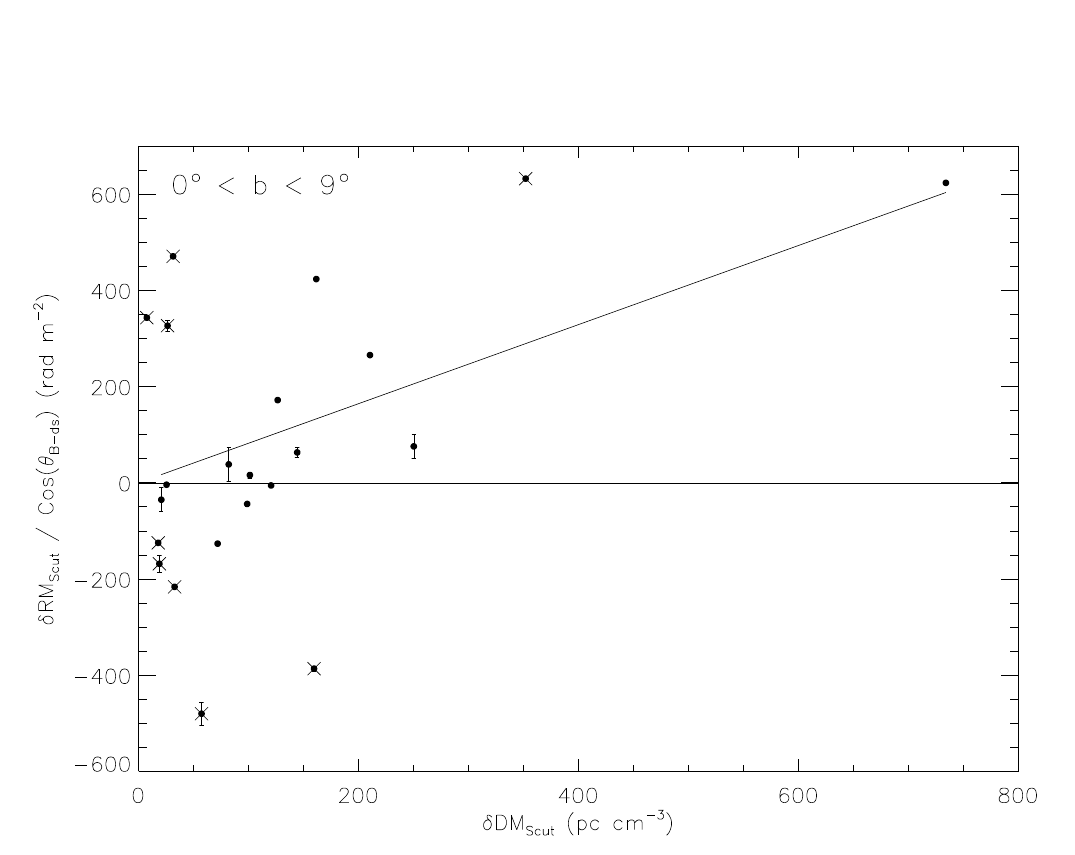} 
\caption{Same as Fig. \ref{fig:LocalRamp}, except for pulsars in the Scutum Arm.
}
\label{fig:ScutRamp}
\end{figure*}

\begin{figure*}
\includegraphics[trim=0.2in 0.6in -0.5in  5in, scale=0.4, width=0.52\linewidth,clip] {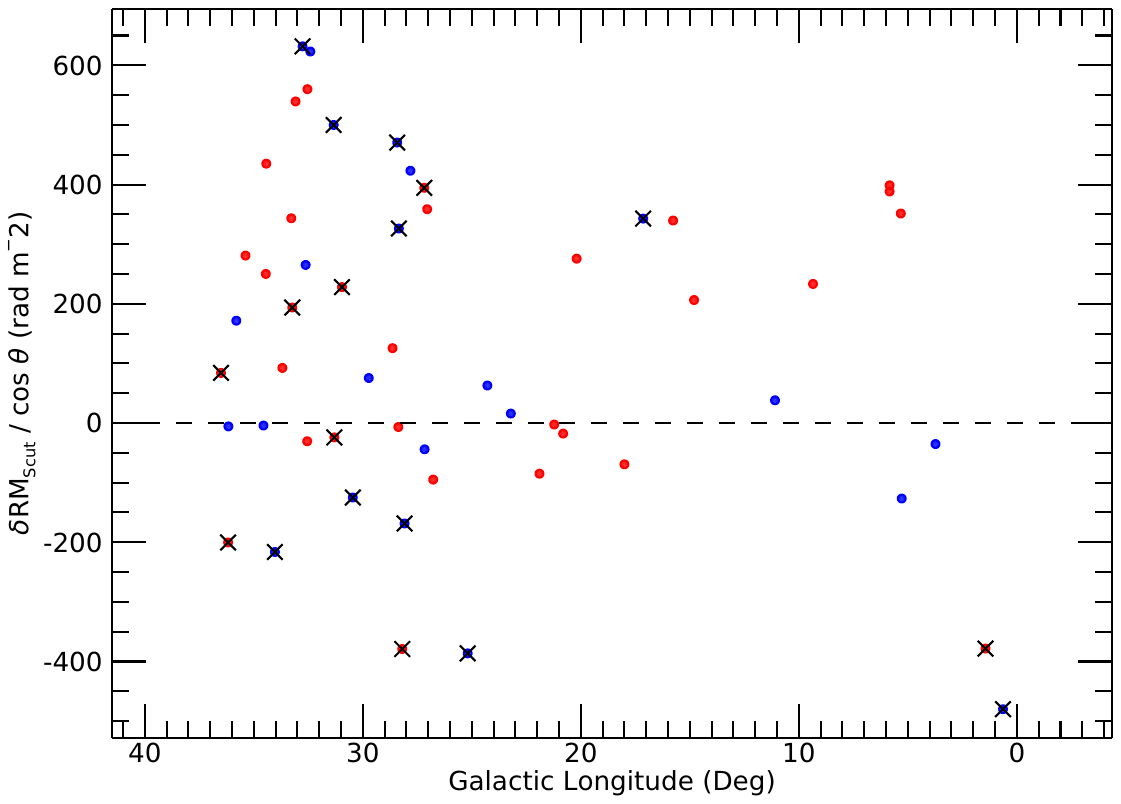} 
\includegraphics[trim=0.2in 0.6in -0.5in  5in, scale=0.4, width=0.52\linewidth,clip]{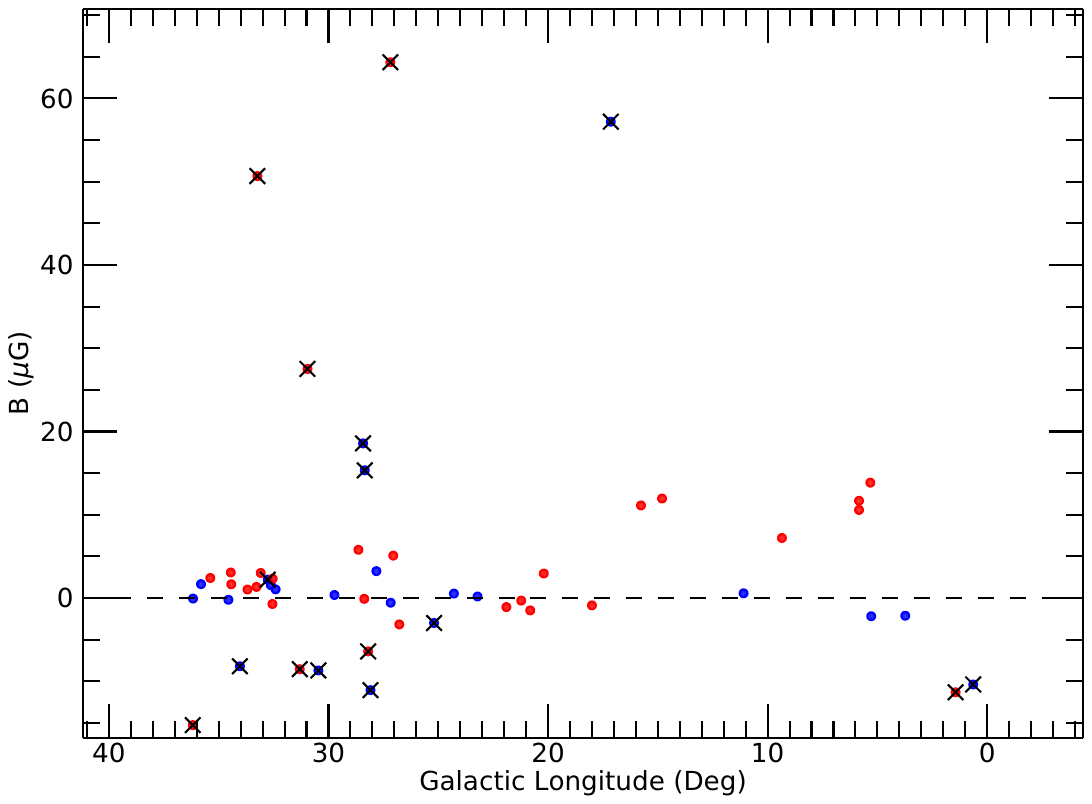} 
\caption{\textit{Left figure:} The geometrically-corrected $\delta $RM$_{\rm Scut}$  as a function of Galactic longitude, for the Scutum Arm pulsars of Fig. \ref{fig:ScutRamp}. \textit{Right figure:} Same as left panel,  except that the magnetic field $B$ is displayed (positive is  CCW) along the Scutum Arm. Pulsar LOS at negative (non-negative) Galactic latitudes are given red (blue) symbols. Outliers are denoted with an $\times$ and listed in Table 2.}
\label{fig:ScutRMBLongi}
\end{figure*}

We study only that part of the Crux-Scutum Arm lying in the first Galactic quadrant, where it is called the Scutum Arm. We define the outer boundary of the Scutum Arm to be identical to the inner boundary of the Sagittarius-to-Scutum Interarm Region, while we set a Galactocentric inner  limit in a manner  similar to our procedure for previous zones.  Next, we reject pulsars lying between these inner and outer spiral boundaries whose pulsar-to-Earth LOS cross the next inward zone. 
The line demarcating  this final rejection criterion is shown in Fig. \ref{fig:Scut} along with our chosen inner and outer Arm boundaries, and the locations and RMs of pulsars lying within all of these boundaries.

For the set of Scutum Arm pulsars shown in Fig. \ref{fig:Scut}, we calculate the geometrically-corrected $\delta$RM$_{\textrm{Scutum}}$ as a function of $\delta$DM$_{\textrm{Scutum}}$ separately for both negative and positive latitude pulsars. We show these quantities for the negative (positive) Galactic latitude pulsars from Fig. \ref{fig:Scut} in the left (right) panel of Fig. \ref{fig:ScutRamp}. For negative (positive) Galactic latitude pulsars, the best-fit slope is $+1.7 \pm 0.3$ ($+0.8 \pm 0.2$), which implies a magnetic field of $+2.1 \pm 0.3$($+1.0 \pm 0.2$) $\mu$G, both pointing in the CCW direction. 

While the negative-latitude fit is significant ($> 6\sigma$), it appears to be dominated by four pulsar LOS at large $\delta \rm{DM}_{\rm Scut}$. However, there are no indications that these four LOS are unrepresentative as they all originate from different longitudes and thus are not all behind a common perturber such as an HII region or magnetic bubble. Furthermore, a fit excluding these four points changes the slope by $<1\sigma$.

The fit for positive-latitude pulsars is again significant  ($> 4\sigma$), 
but it is worth noting that many of the lowest-DM points (i.e., those whose pulsar-Earth LOS originate toward the Earthward side of the Arm) tend to lie below the best-fit line. However, similar to the negative-latitude pulsars, they all originate from a range of longitudes and thus are not all behind a common perturber such as an HII region or magnetic bubble. This suggests that we did not adequately model the magnetic field prior to this Arm for certain pulsar LOS, or that we did not satisfactorily  model its Earthward boundary. Additional  measurements in the future should help to resolve this issue. 

We study the Scutum Arm magnetic fields along {\it{individual}} pulsar LOS in Fig. \ref{fig:ScutRMBLongi}. We do not find any trend with longitude for the positive-latitude LOS. However, negative-latitude LOS have significantly higher $B$ at $0\arcdeg  \lesssim \ell \lesssim 15\arcdeg$ than at $\ell \gtrsim 15\arcdeg$. The average magnetic field strength within $0\arcdeg  \lesssim \ell \lesssim 15\arcdeg$ is $\sim$10 $\mu$G, significantly higher than our average fitted value. 

Previous studies of the magnetic field in the Scutum Arm have found varying magnetic field magnitudes and even opposite directions. Using extragalactic sources and pulsars, \cite{VanEck2011} found a weak CW magnetic field within this region. Later, \cite{Han2018} found a CCW magnetic field of strength 0.4 $\pm$ 0.4 $\mu$G in the Scutum Arm in the first quadrant, a field strength consistent with zero. The more recent work by \citet{Xuet2022} finds a non-zero CCW magnetic field in the Scutum Arm of 1.2 $\pm$ 0.6 $\mu$G, in agreement within 1$\sigma$ with our derived magnetic field both above and below the plane. However, as discussed previously, given that we employ a different technique than \cite{Han2018} and \citet{Xuet2022}, direct comparison is difficult.

\subsection{Perseus-To-Sagittarius Interarm} 
\label{sssec:PersToSag}

As noted in \S\ref{sssec:local}, our Local Arm's low-Galactic-longitude limit lies at the Galactocentric azimuth of the low-longitude end of the YMW17-defined  (red-dashed) arm. We define the Perseus-to-Sagittarius Interarm Region as the extension of the Local Arm toward lower Galactic longitudes, and between the upper boundary of the Sagittarius Arm (see \S\ref{sssec:Sag}) and the lower boundary of the Perseus Arm (see \S\ref{sssec:Per}.) Given the lack of pulsar RMs in our sample within the Perseus-to-Sagittarius Interarm Region, we do not model the magnetic field within it. 

\subsection{Perseus Arm} 
\label{sssec:Per}

The Arecibo telescope could access significant portions of the Perseus Arm in both the first and 
third Galactic quadrants. We also study this Arm in the  second quadrant exclusively via 
others' measurements, thereby linking the two Arecibo-accessible Perseus regions into a single long zone for our work. 

We define the Perseus Arm center via the YMW17 model and its inner boundary to be identical to the outer boundary of the Local Arm, but extending to much higher and lower Galactic longitudes. We define the outer boundary of the Perseus Arm using the same procedure applied to previous arms.
Our Perseus Arm analysis zone is truncated at both ends where pulsar LOS pierce poorly understood intervening regions. The high-longitude limit of our analysis is set so as to avoid any lines of sight sampling the  magnetic field in the Gum Nebula. The low-longitude limit is set to avoid LOS which sample the Perseus-to-Sagittarius Interarm Region  since we were unable to model it (see \S\ref{sssec:PersToSag}). 

\begin{figure}[b]
\includegraphics[trim={1.1in   4.7in 0in  2.1in}, clip,scale=0.52 ]{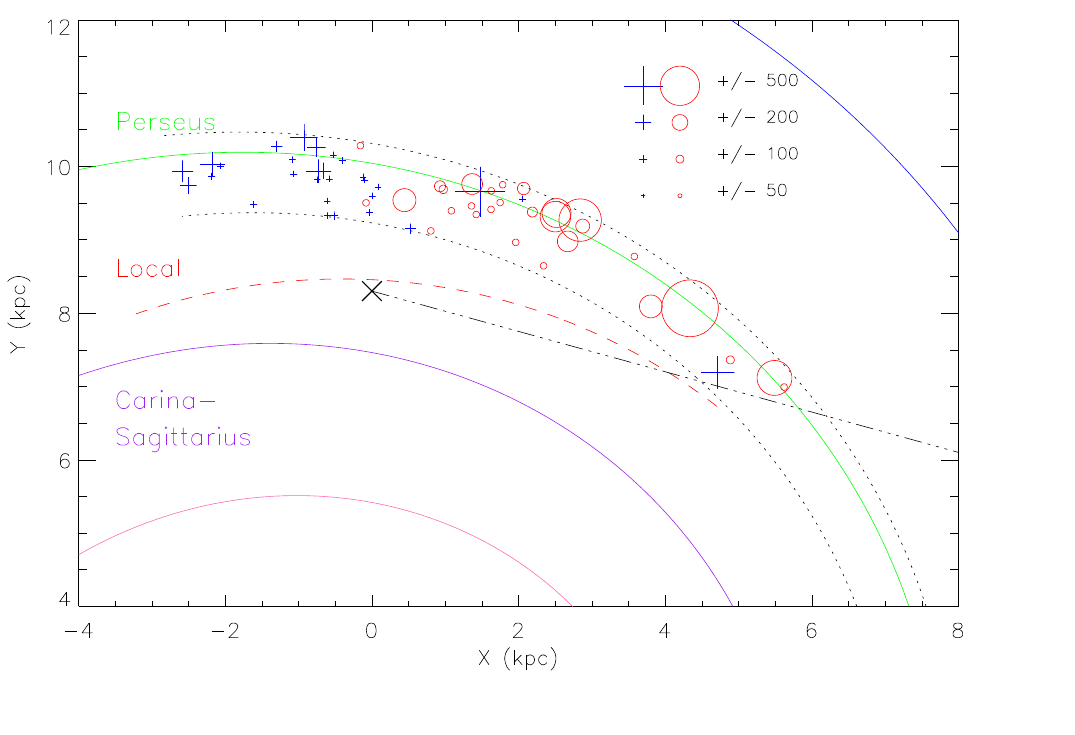}
\caption{
Same as Fig. \ref{fig:Local}, except for the Perseus arm. The dot-dashed line represents the Galactic longitude limit below which the LOS would also sample the Sagittarius-to-Perseus Interarm.}  
\label{fig:PersNoIntArm}
\end{figure}

\begin{figure*}[t!]
\includegraphics[trim=0.95in   5in 0in  1.2in, scale=0.4, width=0.52\linewidth,clip ]{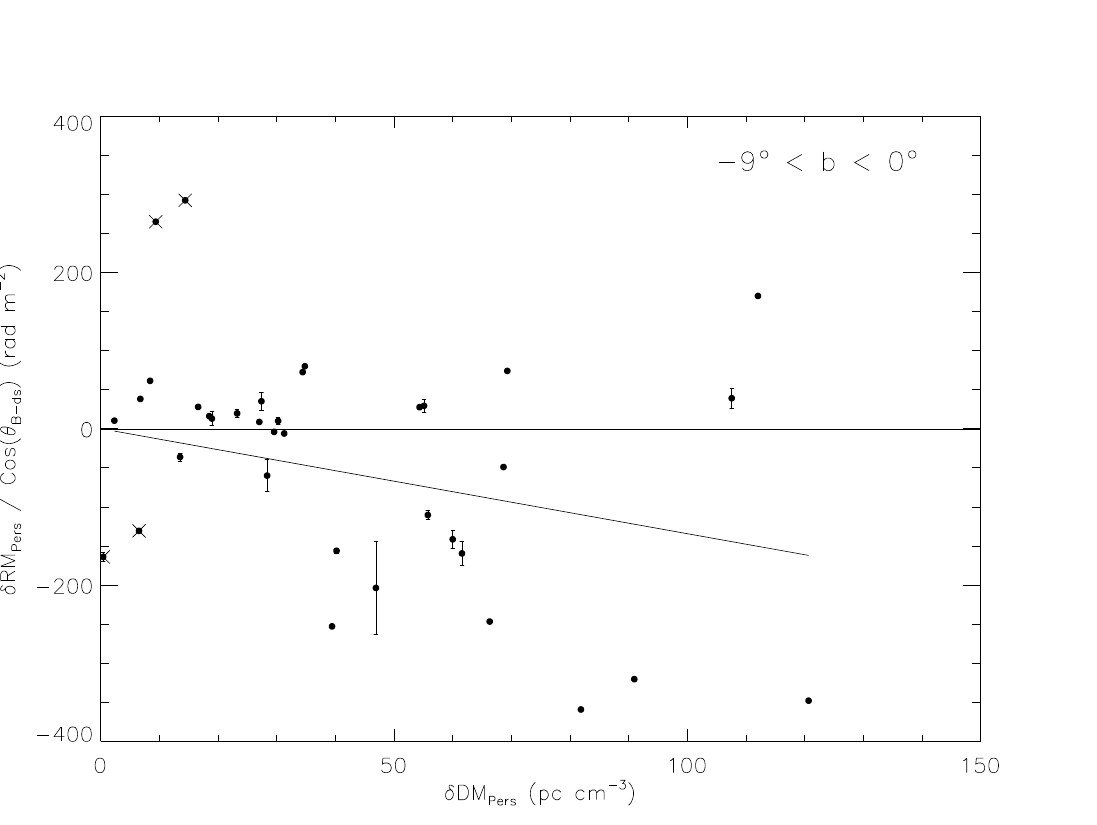} 
\includegraphics[trim=0.95in   5in 0in  1.2in, scale=0.4, width=0.52\linewidth, clip ]{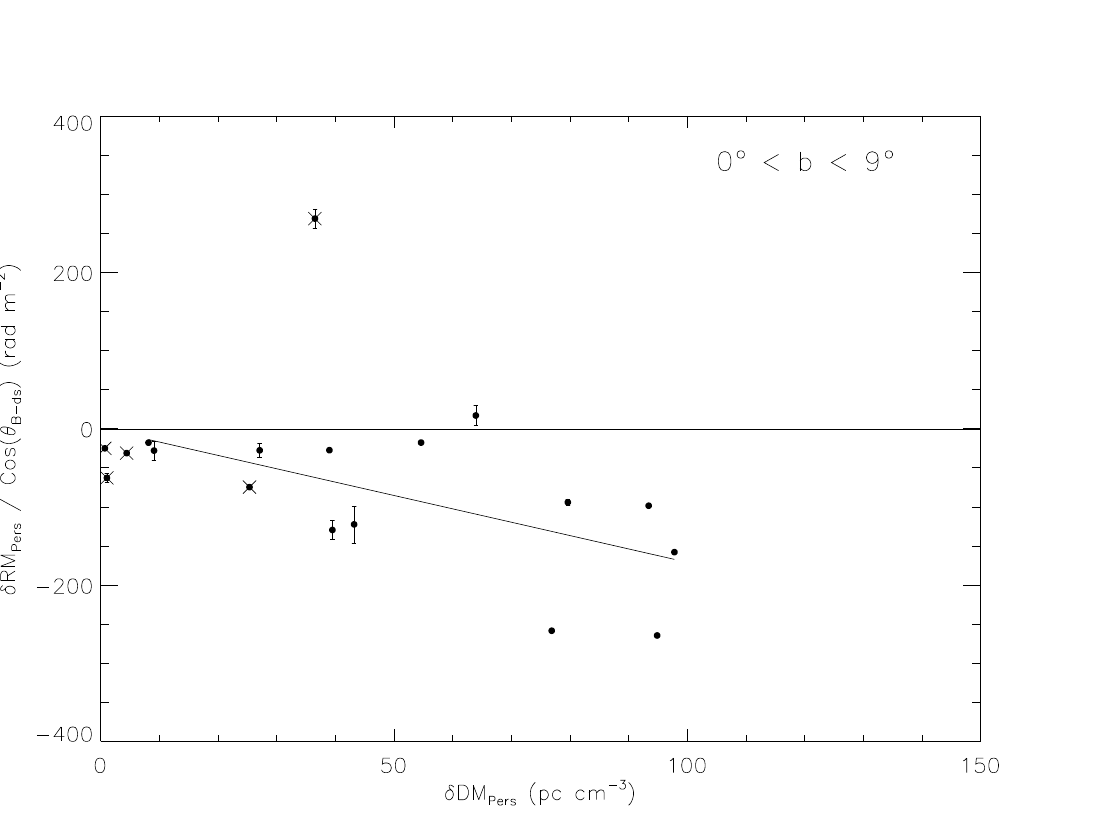} 
\caption{Same as Fig. \ref{fig:LocalRamp} except for pulsars in the Perseus Arm. We truncate the right plot at $\delta$RM$_{\rm Pers}$ = [$-500$, $500$] rad m$^{-2}$ and hence two outliers at ($\delta$RM$_{\rm Pers}$,  $\delta$DM$_{\rm Pers}$) = (735.79 rad m$^{-2}$, 63.24 pc cm$^{-3}$) and at ($-1267.22$ rad m$^{-2}$, 17.41 pc cm$^{-3}$) are not shown. We also truncate the plot at $\delta$DM$_{\rm Pers} < 150$ pc cm$^{-3}$, with an outlier at ($\delta$RM$_{\rm Pers}$,  $\delta$DM$_{\rm Pers}$) = (-307.25 rad m$^{-2}$, 349.79 pc cm$^{-3}$) not shown. All outlier information (including these two) is listed in Table  \ref{table:outliers}.}
\label{fig:PersNoIntArmRamp}
\end{figure*}

All pulsars within our  Perseus Arm analysis zone are shown in Fig. \ref{fig:PersNoIntArm}. 
For the set of Perseus Arm pulsars shown in Fig. \ref{fig:PersNoIntArm}, 
we calculate the geometrically-corrected $\delta$RM$_{\textrm{Perseus}}$ as a function of 
$\delta$DM$_{\textrm{Perseus}}$ separately for both negative and positive latitude pulsars 
using Eqs. \ref{eqn:deltaRMbroke} and \ref{eqn:deltaDMbroke}. We show these quantities for 
the negative (positive) Galactic latitude pulsars from Fig. \ref{fig:PersNoIntArm} in the 
left (right) panel of Fig.  \ref{fig:PersNoIntArmRamp}.  

For negative (positive) latitude pulsars, the best-fit slope is $-1.4 \pm 0.6$ ($-1.7 \pm 0.4$), which implies a magnetic field of $-1.7 \pm 0.7$ ($-2.1 \pm 0.6$)  $\mu$G, both pointing in the CW direction along the Arm\footnote{As noted in \S\ref{sssec:analysis}, the Perseus Arm is the only zone analyzed whose covariance between two arms' fitted slopes (in this case Local and Perseus) leads to a significant enhancement of the total uncertainty.}. The fit for negative-latitude pulsars has  a significance of 2.4 $\sigma$ with notable scatter about $\delta$RM$_{\rm Pers} \sim 0$. The fit for  positive-latitude pulsars is more significant (3.8$\sigma$) with relatively less scatter about $\delta$RM$_{\rm Pers} \sim 0$. For both the  negative and positive latitude fits, we do not find any additional evidence for a longitude or latitude dependence of the magnetic field other than that mentioned above. 

In addition to the Perseus Arm regions excluded from analysis as described above, our following discussion of the Perseus Arm field must be qualified by the fact that LOS toward it intersect a less-completely sampled part of the  Local Arm's magnetic field, namely those parts lying toward larger galactocentric radii (see  \S\ref{sssec:local} and its Fig. \ref{fig:Local}). Hence our estimates of the intervening Local Arm field may not be  fully representative of the LOS toward the Perseus Arm. It will be possible
to remedy this issue in the future, as more RMs in the outer regions of the Local Arm become available.

A CW magnetic field within the Perseus Arm is similar to that found by \cite{Mitra2003} and \cite{VanEck2011}. Using RMs from 11 pulsars along the direction of the Perseus Arm, \cite{Mitra2003} determine a CW magnetic field with magnitude $-1.7 \pm 1.0$ $\mu$G. This is in agreement with both our negative and positive Galactic latitude magnetic field derivations for the Perseus Arm. \cite{VanEck2011} use EGS measurements and also find a CW magnetic field for the Perseus Arm. However, \cite{VanEck2011} find that while the magnetic field tends to follow the spiral arms in the inner Galaxy, it is largely azimuthal in the outer Galaxy (where the outer Galaxy is defined as that part lying outside of the Local Arm). Thus, \citet{VanEck2011} do not provide a field strength for the Perseus Arm alone. We note that the Perseus Arm is largely contained within region 5C of their Fig. 6 for which the magnetic field strength is $-0.86 \pm 0.09 $ at $R=8.5$kpc and decreases as a function of $R^{-1}$. 

Unlike the CW field determination of \cite{Mitra2003}, \cite{VanEck2011}, and the current work, \cite{Wetal2004} and \cite{Han2018} both attribute a CCW magnetic field to the Perseus Arm. \cite{Wetal2004}  model the Perseus Arm using a small sample of pulsars within the large longitude range of 60$\degree$ to 78$\degree$. While their sample is limited to eight pulsars, they find two field reversals in this longitude region: one at ${\rm d}\sim 4.5 \pm 1$ kpc and  ${\rm d}\gtrsim 6$ kpc  (see their Fig. 9). Similarly, \cite{Han2018} find an increase in pulsar RMs for distances $>5$ kpc in the longitude region of $60 \degree < l < 80 \degree$. However, the longitude region  used in both  \cite{Wetal2004} and \cite{Han2018}  includes the poorly understood Perseus-to-Sagittarius Interarm (see \S\ref{sssec:PersToSag}) and hence the field reversal could also be due to a CCW magnetic field within the Interarm. These authors similarly do not apply any geometrical corrections.

\cite{Han2018} also use the RMs of EGS to deduce a switch in the magnetic field direction between the Local Arm and the location of EGS in the longitude interval of $80 \degree < l < 120 \degree$.  They locate this switch to the Perseus Arm, although it cannot be definitively located with EGS. Using our updated sample of pulsars, we model the same wedge region of $80 \degree < l < 120 \degree$ and do not find any field reversal within the Perseus Arm, so the reversal must lie somewhere beyond it.

\subsection{Norma-Outer Arm and Perseus-to-Norma Interarm} \label{sssec:normaOuter}

\citet{Rankin2023} greatly increased the number of pulsars with RMs in the  Perseus-to-Norma Interarm and the Norma-Outer Arm  e.g., see Fig. \ref{fig:AreciboFace}. However, even with this increase, there are still an insufficient number of pulsars within these two regions to do  proper studies of the magnetic field. With the increasing sensitivity of radio telescopes, though, more pulsars will undoubtedly be found within these regions, and our technique can then be applied to them.

\section{Comparison Of Our Results with Galactic Magnetic Field Models}
\label{sssec:Discuss} 

Mapping the magnetic field structure within galaxies (both our own and external ones) is important for understanding how galaxies form and evolve. The magnetic structure affects processes such as cosmic ray propagation, the outflow of gas, and the star formation rate \citep{Beck2016}. Additionally, understanding the magnetic field structures of spiral galaxies gives essential insight into the formation of the Milky Way. 

\subsection{Observations and Theories of Spiral Galaxy Magnetic Fields}
\label{sec:extragalobs}
Observations of external spiral galaxies have revealed that the magnetic field typically follows the spiral structure, with the magnetic field pitch angle similar to that of the optical spiral arms \citep{Beck2016, Han2017, Krause2019} and the magnetic field strengths peaking in the interarm regions (but not the arms) e.g., see NGC 6946 \citep{Heald2009}. Note, however,
that the LOS probed in external galaxies typically extend to much larger $|z|$ than do Milky Way pulsar observations, so the extragalactic results are not directly comparable with our Galactic measurements.

The magnetic structure within spiral galaxies is most often understood under mean-field dynamo theory. In this theory, magnetic fields are generated and maintained by the differential rotation of plasma within the disk of the Galaxy along with turbulent motion due to supernovae, stellar winds, cosmic rays, etc. (referred to as the $\alpha$-effect). The differential rotation continuously generates a toroidal field, while the 
$\alpha$-effect regenerates the poloidal field, creating a self-sustaining magnetic field configuration \citep[See Chapter 5.5 of][ for further details on dynamo theory]{Kronberg2016}. While a spiral magnetic field structure is supported by dynamo theory, many of the observed details of extragalactic spiral galaxies' magnetic structure are not.  For example, extragalactic observations of a similar pitch angle between the optical and magnetic arms along with a peak in field strength in the interarm regions are not well explained by the dynamo model \citep{Beck2016, Krause2019}. 

Under mean-field dynamo theory, different modes are possible for the azimuthal symmetries of the magnetic field within the disk of a spiral galaxy \citep{Beck2016}. The most common dynamo mode is one in which the magnetic field in the disk is an axisymmetric spiral. M31 has such a field \citep{Beck2016}. However, a dynamo model could also support a bisymmetric spiral magnetic field configuration in which there is a single reversal of the field direction at a given disk radius \citep[see Fig. 5.7 in][for further discussion of the difference between an axisymmetric and bisymmetric magnetic field configuration]{Kronberg2016}. A bisymmetric magnetic field configuration within a galaxy has yet to be conclusively confirmed, although it is posited that M81 may have such a structure \citep{Beck2016}. 

\subsection{Past Efforts to Model the Galactic Magnetic Field}
\label{subsec: past efforts}

There have been multiple efforts to use various types of observations to constrain models of the Galactic magnetic field. The most common models advanced include  an axisymmetric spiral, (hereafter, ``ASS'') model, a bisymmetric spiral, (hereafter, ``BSS'') model, and an ASS + ring model. In the latter model, the ASS field is over laid by a field reversal within a narrow ring at a given Galactic radius. Below, in order to give context to our efforts, we give brief overviews of some of the recent efforts to fit these models to pulsar and/or EGS RMs plus additional types of data in some cases. We also list the relevant details of each model in Table \ref{table:FieldModels}. 

\begin{deluxetable*}{lll}
\tablecolumns{3}
\tabletypesize{\small}
\tablecaption{Overview of Relevant Galactic Magnetic Field Models \label{table:FieldModels}}
\tablehead{
\colhead{Reference} & \colhead{Input Data\tablenotemark{a}} & \colhead{Notable Results\tablenotemark{b}} 
}
\startdata
\cite{Sunet2008} & EGS RMs; All-sky & \textit{Disk field:} CW ASS field with a CCW ring between the \\ 
& Intensity \& Polarization Maps & Carina-Sagittarius Arm and the Crux-Scutum Arm. \\  
& & \textit{Toroidal halo field}: odd z-parity\\
\hline
\cite{Men2008} & Pulsar RMs & \textit{Disk field}: Neither ASS, BSS, nor ASS + ring models fit the data. \\  
\hline
\citet{VanEck2011} & EGS RMs; Pulsar RMs & \textit{Disk field:} CW ASS field in inner galaxy with a single CCW  \\
& & spiral interior to the local arm. CW ring in outer galaxy.  \\
\hline
\citet{Jansson2012} & WMAP7 Galactic synchrotron & \textit{Disk field}: One azimuthal ring between 3 to 5 kpc; Eight  \\ 
& emission; EGS RMs  & spiral arms with multiple field reversals \\ 
& &  \textit{Toroidal halo field}: Field has z-asymmetry \\
\hline
\citet{Han2018} & Pulsar RMs & \textit{Disk field:} Field follows the spiral arms with field reversals   \\ 
&  & at the boundaries of some arms and interarms \\ 
\hline
\citet{Xuet2022} & Pulsar RMs &  \textit{Disk field:} Similar to  \citet{Han2018} \\ 
\hline
\citet{Dickeyet2022} & GMIMS Faraday depth spectra & \textit{Disk field:} Spiral or azimuthal disk field \\
& of the diffuse Galactic synchrotron & \textit{Toroidal halo field:} Field has z-asymmetry\\
& emission; EGS RMs & \\
\hline
\citet{Unger2023} & WMAP \& Planck Galactic & \textit{Disk:} Grand-design spiral with alternating-sign magnetic arms, \\ 
& synchrotron emission; & with magnitude anticorrelated with other spiral tracers.  \\ 
& EGS RMs& Alternatively, Local Spur Model fits adequately. \\  
& & \textit{Toroidal halo field}: odd z-parity\\
\hline
\citet{XuHan2024} & EGS RMs; Pulsar RMs  & \textit{Toroidal halo field}: odd z-parity\\
\enddata 
\tablenotetext{a}{Type of data used to perform the Galactic magnetic field analyses}
\tablenotetext{b}{Brief summary of relevant model results.}
\end{deluxetable*}

\subsubsection{Axisymmetric and Axisymmetric + Ring Models}

We first focus on  observations and analyses  that appear to be best described by an ASS spiral Galactic magnetic field with or without a ring. \cite{Sunet2008} use the RMs from EGS along with all-sky intensity and polarization maps to study the Galactic magnetic field. They find that the best model comprises a CW, ASS spiral magnetic field atop most arms, with a CCW magnetic ring generally between the Carina-Sagittarius and  Crux-Scutum Arms. This result is similar to that of \citet{VanEck2011} who, using the RMs of EGS, find that the inner\footnote{For our purposes, the ``inner Galaxy'' refers to that portion lying within the Solar Circle.} Galactic  magnetic field points largely CW along spiral arms, except for a single CCW spiral field interior to the Local Arm. In the outer galaxy, the field is consistent with a CW azimuthal field. \citep[see Fig. 11 of][]{VanEck2011}. This model is similar to the  model of \cite{Sunet2008}, which consists of a CW ASS field with a single CCW ring near the Sagittarius arm. Interestingly, the major magnetic spiral feature in the  \citet{VanEck2011}  model overlays the generally accepted Carina-Sagittarius Arm location in the first Galactic quadrant, but not in the other three quadrants.

\subsubsection{Higher-Order models}
We next focus on observations that are not well-fit by the simpler models described above. For example, \cite{Men2008} use pulsar RMs and find that neither an ASS, BSS, nor ASS + ring model can adequately fit the measurements. Instead, they conclude that a more complex model is needed. \citet{Han2018} similarly use pulsar RMs and find that the field follows the spiral arms with field reversals at the boundaries of some arms and interarms. \citet{Xuet2022} find similar results using the same technique as \citet{Han2018} but with a larger sample of pulsar RMs. 

\citet{Unger2023} used RMs from EGS to fit for different grand spiral configurations. Their best-fitting model has 6 magnetic arms with field magnitudes peaking in the interarms with near-zero field strength at the arm centers. This is consistent with some observations of extragalactic spirals (see \S\ref{sec:extragalobs}), but
not with most Galactic models. However, the RMs used by \citet{Unger2023} are also fit approximately equally well with a model in which the magnetic field is zero everywhere except the Local Arm, where it is strong ($\sim 4 \mu$G) and CCW. This demonstrates a major difficulty in using solely EGS RMs to model the Galactic magnetic field, as they do not explicitly encode Galactic distance information in the same way as pulsars.

\subsubsection{Magnetic Field Disparities Above and Below the Galactic Plane}
\label{sec:UpDown}

\cite{Sunet2008}, \citet{Jansson2012}, \citet{Dickeyet2022}, \citet{Unger2023}, and \citet{XuHan2024} have all found evidence for a toroidal (i.e., longitudinal) halo field that extends to high $|z|$ and has different properties above and below the Galactic plane.  In particular, all of the above authors find that the toroidal halo field has opposite signs above and below the Galactic equator. At low $|z|$, the Sagittarius Arm's longitudinal field also exhibits an asymmetry between positive and negative latitudes, as
was previously noted (see \citet{Ordoget2017}, \citet{Maet2020}, and our \S\ref{sssec:Sag}). 

\subsection{Our Work in Context}
\label{subsec: our work}
\begin{figure*}[]
{
	\includegraphics[trim=1.5in 5in 2in  1in, scale=0.4, width=0.49\linewidth ]{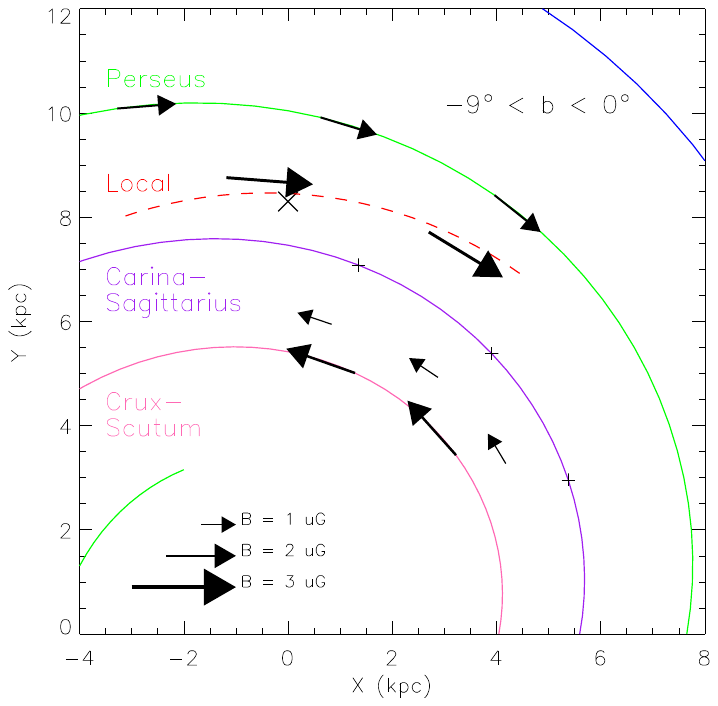} 
	\includegraphics[trim=1.5in  5in 2in 1in, scale=0.4, width=0.49\linewidth   ]{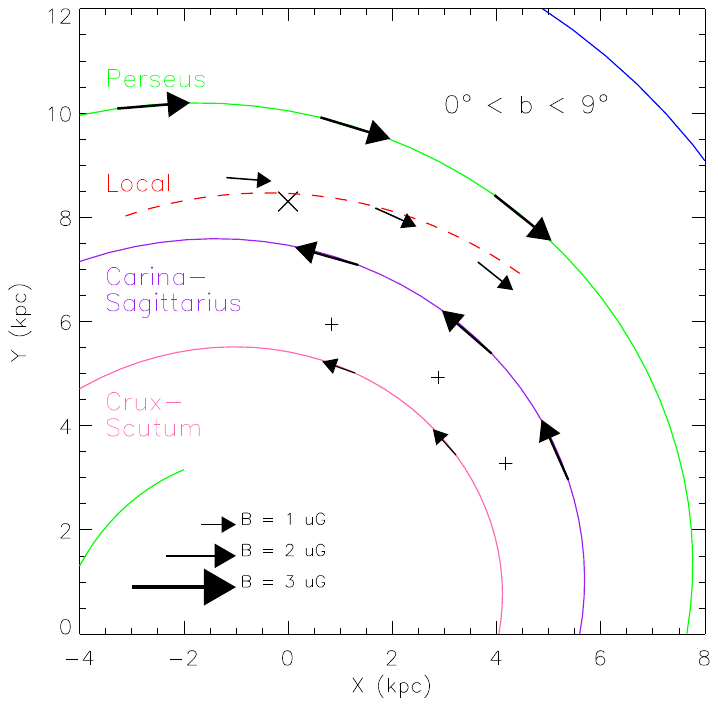} 
}
\caption{\textit{Left figure:} The derived magnitudes and directions of the large-scale, planar magnetic field for negative Galactic latitudes. The magnetic field vectors  are displayed atop the centers of the associated arms, except for the Sagittarius-to-Scutum interarm field which is shown between the Sagittarius and Scutum arms. Arrow sizes are directly proportional to the the magnitude of the magnetic field, while the arrow directions indicate whether the derived field is CW or CCW. Zones with derived magnetic field strengths less than their 1$\sigma$ uncertainty are shown with crosses. See  Table~\ref{table:Bs} 
for exact magnitudes and their uncertainties. 
\textit{Right figure:} Same as left figure except for positive Galactic latitudes. } 
\label{fig:BFieldSpiralArms}
\end{figure*}

We now discuss our results in the context of the above Galactic magnetic field models. To summarize our work,  we show our best-fit low-latitude magnetic field directions and magnitudes in our zones of study in Fig. \ref{fig:BFieldSpiralArms}, with negative  (positive) Galactic latitudes plotted separately in the left (right) panel.

\subsubsection{Measured Field Strengths Above and Below the Galactic mid-plane}
\label{sssec:fieldabovebelowplane}

Our measured differences in the {\it{disk}} field strength (though not its direction) at positive and negative Galactic latitudes in several zones could be a result of combining the disk's stronger z-symmetric longitudinal field with the {\it{halo's}} weaker z-asymmetric or odd z-parity toroidal fields if these toroidal fields extend down into the Galactic disk (See \S\ref{sec:UpDown} and Table \ref{table:FieldModels}). If these toroidal fields are indeed significant in the Galactic disk, then we would expect our measured B-fields above the plane to tend more positive (i.e., CCW) than those below the plane. The models,
however, suggest that the toroidal halo fields decline sharply toward the plane.  Additionally, we do not see {\it{consistent}} evidence for the asymmetry in all regions.

\subsubsection{Field Reversal in the Disk within the Solar Circle}
\label{sssec: field reversal sag arm}

\begin{figure*}[t!]
\includegraphics[trim=1.2in  5in 1.5in  1.9in, scale=0.4, width=0.52\linewidth, clip]{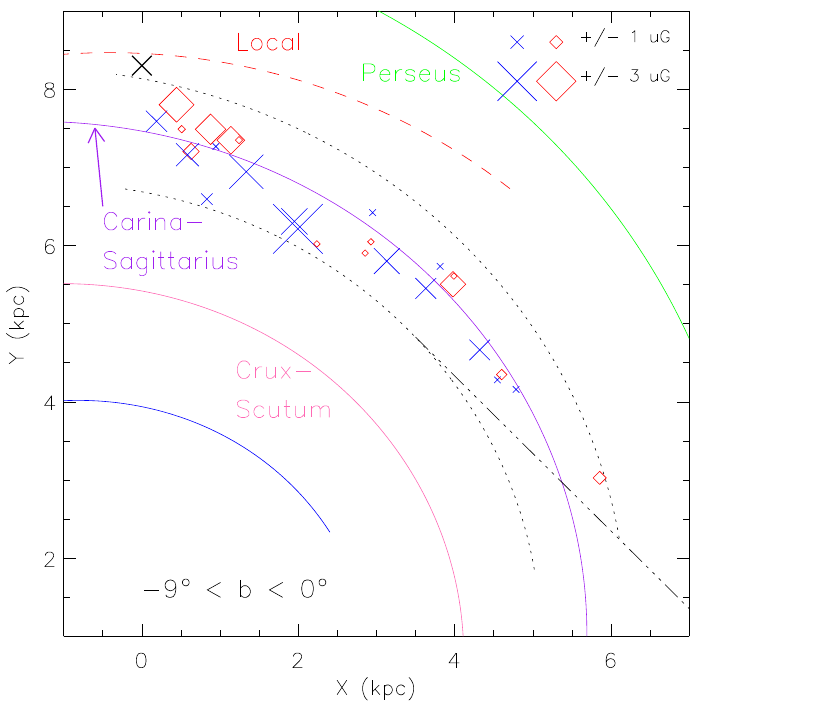} 
\includegraphics[trim=1.2in   5in 1.5in  1.9in, scale=0.4, width=0.52\linewidth, clip ]{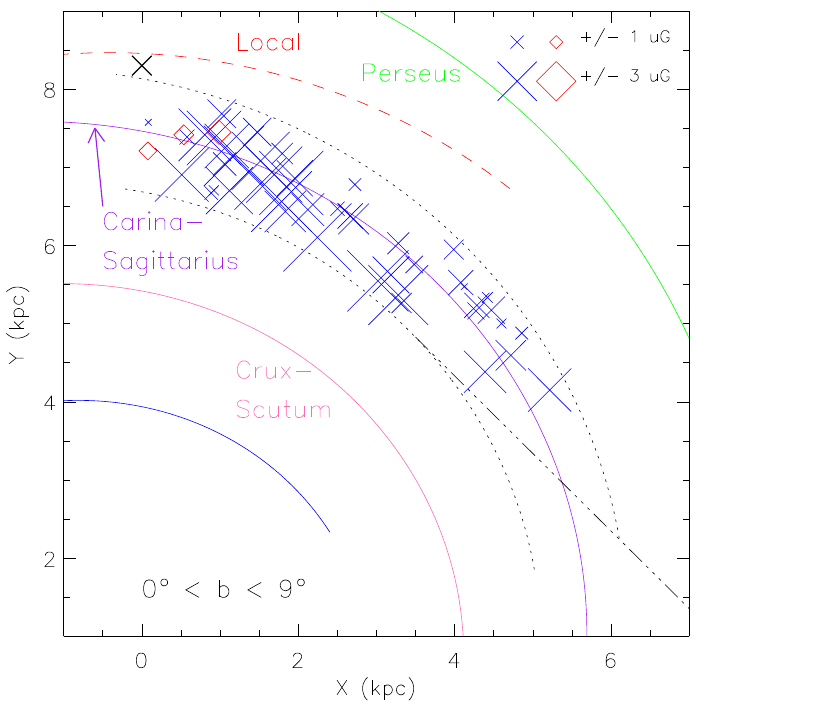} 
\caption{\textit{Left figure:} Longitudinal magnetic fields at negative Galactic latitudes in the Sagittarius Arm, derived from pulsars at the displayed locations. The Sagittarius Arm Center is delineated and labeled in purple with its inner and outer boundaries are shown by flanking black dotted lines. A CCW magnetic field (``positive'' by our convention) is shown as a blue $\times$, while a CW magnetic field (``negative'' by our convention) is indicated with a red diamond. The Solar System is marked by the black $\times$ at (x, y) = (0.0, 8.3) kpc.  Outliers (listed in Table \ref{table:outliers}) are not shown. \textit{Right figure: } Same as left panel, except at positive Galactic latitudes.}
\label{fig:SagBNorthSouth}
\end{figure*}

Both above and below the Galactic plane, we find evidence for only a single field reversal throughout all the Galactic regions we studied, lying at or near the Sagittarius / Local Arm boundary. In Fig. \ref{fig:SagBNorthSouth}, we show our derived Sagittarius arm B-fields along specific pulsar LOS, with negative Galactic latitude pulsars shown in the left panel and positive Galactic latitude pulsars shown in the right. At positive Galactic latitudes, the strength and direction (CCW) of the B-field in the Sagittarius Arm are relatively constant, indicating that the field reversal occurs between the Local (CW) and Sagittarius (CCW) Arms.  However, at negative Galactic latitudes, the CW to CCW field reversal occurs inside of the Sagittarius arm {\it{itself}}. Therefore the reversal apparently conforms to spiral structure in the plane, but tilts toward (away from) the Galactic Center South (North) of the Galactic Plane.

Studying the total RMs of EGS and pulsars, \citet{Ordoget2017} also found evidence of  a dependence on Galactic latitude and longitude for the field reversal from a CW to a CCW field near the Sagittarius Arm. They note that the signs of the total RMs in the region bounded by $56\arcdeg \lesssim l \lesssim 67\arcdeg$ and $-3\arcdeg < b < 5\arcdeg$\footnote{This corresponds to the approximate longitude range of the tangent to the Sagittarius arm.} are mostly separated on the sky by a diagonally-tilted line (hereafter referred to as the Ordog et al. line) extending from $(l, b) = (56\arcdeg, -2\arcdeg$) to $(l, b) = (67\arcdeg, 4\arcdeg$).  However, their EGS and Galactic synchrotron RM measurements do not yield  the distance to the sampled objects, so (as they note), their ``line'' might be merely the projection onto the Plane of the Sky of a more complicated topology such as a tilted planar object.  It is also important to note that
the ``line'' is not a physical object but merely an imaginary  boundary separating RMs of different signs.  

\begin{figure}[b]
\includegraphics[trim={1in   5in 0in  1.2in}, clip,scale=0.5 ]  
{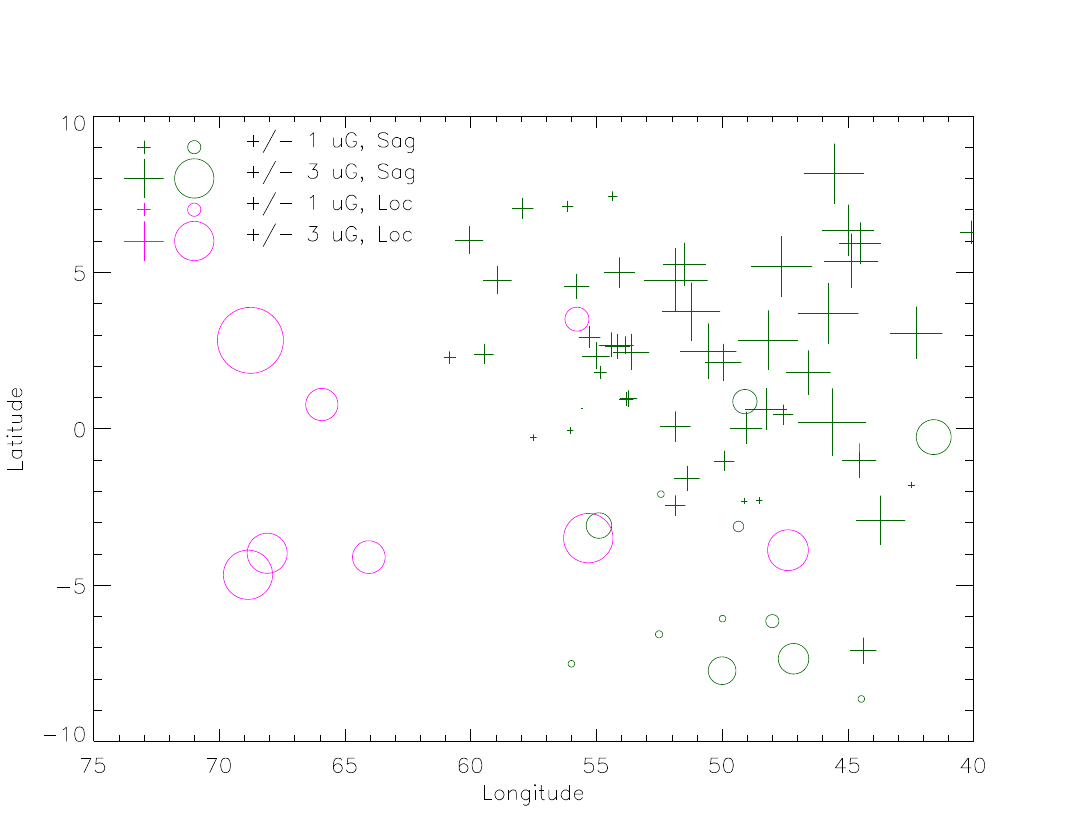}
\caption{Magnetic field strengths measured in this work along the Local (Magenta) and the Sagittarius (Green) Arms in the approximate longitude and latitude regions studied by \citet{Ordoget2017}.  Magnetic field direction along an Arm is indicated with a cross (positive, CCW) or a circle (negative, CW), with its magnitude shown via the symbol's size. For pulsars for which $|\textrm{B}| < 0.5$ $\mu$G, we set the symbol size to be equivalent to that of $|\textrm{B}|=0.5$ $\mu$G. Similarly, for any pulsars for which $|\textrm{B}| > 20$ $\mu$G, we set the symbol size to be equivalent to that of $|\textrm{B}|=20$ $\mu$G. Outliers (listed in Table \ref{table:outliers}) are not shown.}
\label{fig:LongiLatiOrdogZone}
\end{figure}

In contrast to \citet{Ordoget2017}, our analysis techniques separate RMs and B-fields by arm, thereby yielding information on distance, a
spatial dimension that has been mostly absent in previous analyses. In Fig. \ref{fig:LongiLatiOrdogZone}, we show our derived Local and Sagittarius Arm B-fields in approximately the same longitude and latitude region as that studied by \citet{Ordoget2017}.\footnote{We extend their longitude range to include a larger portion of the Sagittarius arm.}  
Consistent with their results, we find that the CW ``negative'' fields are predominantly found toward higher $l$ and lower $b$ (i.e., below the Ordog et al. line) while the CCW ``positive'' fields are primarily located toward lower $l$ and higher $b$ (i.e., above the Ordog et al. line). Given our ability to distinguish between the Local and Sagittarius Arms, we note that the majority of the trends discussed above are naturally accounted for by our and others' results of a CW field in the Local Arm switching to a CCW field in the Sagittarius Arm.  

It is interesting to note that the few Sagittarius Arm LOS with CW ``negative'' B-fields primarily lie below the Ordog et al. line, which is unlike the positive Sagittarius B-field LOS located above the line. Additionally, as discussed above, we find a latitude-dependent location of the field reversal near the Local and Sagittarius Arms. If the Ordog et al. line were in a quasi-planar object that follows the shape of the Sagittarius/Local arm boundary and is tilted toward the Galactic Center at negative Galactic Latitudes, it could explain both our above results and those of \citet{Ordoget2017}, as it would predict that the reversal at negative  Galactic latitudes would occur at smaller  Galactic radii than the reversal at positive latitudes. However, given the complexity of this region and the possible underlying progenitors of the magnetic field, we leave further investigation of this to future work.

\subsubsection{Comparison to External Spiral Galaxy Magnetic Fields}
\label{sssec: comparison observed to other galaxies} 

In both the best-fit Galactic model by \citet{Unger2023} and in many observations of external spiral galaxies \citep{Heald2009, Beck2016}, the magnetic field strength peaks {\it{between}} arms, with negligible field strengths {\it{within}} them. Our negligible field strength measurements in the negative Galactic latitude Sagittarius Arm  thus agree both with the \citet{Unger2023} model and with extragalactic  spiral galaxies' magnetic field observations. However, no other zone that we analyzed is consistent with this picture. Notably, the magnetic field within the one distinctive interarm region that we studied, the Sagittarius-to-Scutum Interarm Region, is consistent with zero at negative Galactic latitudes and relatively weak at positive Galactic latitudes.

\subsubsection{Bisymmetric Model}
\label{sssec: bisymmetric model}
Both above and below the plane, our RM values support a bisymmetric model that includes at least one magnetic field reversal. It is possible that there are more magnetic field reversals interior to the Scutum Arm or beyond the Perseus Arm, as these zones are not modeled in this work. A bisymmetric model is in general agreement with recent efforts to study the Galactic magnetic field in the disk; e.g., \citet{Han2018, Xuet2022, Unger2023}. However, the field reversals and field directions derived in this work disagree with many of those previously published. We attribute this to different analysis procedures; particularly our improved technique which accounts for intervening arms' RM contributions and the dot-product nature of Faraday rotation.

\section{Summary}
\label{sssec:Concl}

In this work, we develop a new technique for using the observed total RMs of an ensemble of pulsars to study the large-scale Galactic magnetic field under the assumption that the  field follows the spiral arms. We split the observed total RM and DM of a given pulsar into segments corresponding to different spiral arms or interarm zones along a given LOS. Starting with the Local Arm, we solve for its magnetic field strength while accounting for the dot-product nature of the RM-defining integral. We then recursively move to zones farther from Earth, using our calculation of the inner zones' magnetic field strengths to determine and subtract their $\delta$RM contributions to the total RM, leaving only the outer most zone's contribution. In this fashion, we are able to assemble an arm-by-arm picture of the Galactic magnetic field within several kpc of the Earth.

We apply our new technique to 313 low-Galactic latitude pulsar RMs published in \citet{Rankin2023} along with other pulsar RMs within and adjoining the same regions (approximately the first Galactic quadrant and a section of the third Galactic quadrant). Our technique adds approximate distance information to otherwise two-dimensional RM analyses. Its prime application is in separating spiral arms along a line of sight for individual study, but we also show its utility in providing likely explanations for puzzling features in the total-RM sky. 

We determine the magnetic field strength along the Local Arm, the Sagittarius Arm, the Sagittarius-to-Scutum Interarm, and the Perseus arm. We find disparities $>1\sigma$ for the magnetic field strength (though not its direction) above and below the plane in most zones studied, with the greatest disparity in the Sagittarius Arm. 

We find only one major Arm-to-Arm (or Arm-to-Interarm) field reversal throughout all the zones we investigate, although its Galactocentric location varies slightly between positive and negative Galactic latitudes. At positive $b$, the field reversal occurs between the Local (CW) and Sagittarius (CCW) Arms, while at negative $b$, the reversal appears to occur within the Sagittarius Arm itself.

Our results  favor a bisymmetric model for the large-scale Galactic magnetic field with at least one field reversal inside the solar circle. It is possible that there are more reversals beyond our current region of study, either inside the Scutum Arm or outside the Perseus Arm, but we leave analysis of these zones to future work. A bisymmetric model is in agreement with most recently published works on the Galactic magnetic field \citep{Han2018, Xuet2022, Unger2023}. However, unlike observations of external spiral galaxies and certain Galactic models, we find that the magnetic field is usually strongest \textit{within} arms rather than \textit{between} then.

We expect that the geometrically-corrected, `arm-by-arm' technique presented here will continue to be useful as additional pulsars' RMs and DMs are measured. Future work could then extend this analysis to farther regions of the Galaxy. 

\begin{deluxetable*}{lccccc}
\tablecolumns{8}
\tablecaption{Outlier $\delta$RM/$\cos \theta$, $\delta$DM, and $B$  thresholds by zone}
\label{table:thresholds}

\tablehead{
\colhead{Fitted Zone}   
&\multicolumn{2}{c}{$\delta\rm RM_{zone}/\cos\theta$  Outlier }    
& \colhead{$\delta\rm DM_{zone}$ Outlier}      
& \multicolumn{2}{c}{$B$ Outlier}    \\
\colhead{}                
&\multicolumn{2}{c}{ Threshold (rad m$^{-2}$)} 
& \colhead{ Threshold (pc cm$^{-3}$)}   
& \multicolumn{2}{c}{ Threshold ($\mu$G)} \\
\hline
 \colhead{} & 
 \colhead{ Lower } 
 &  \colhead{ Upper} 
 &  \colhead{ Upper } 
 &  \colhead{ Lower}  
 &  \colhead{Upper}          
}
\startdata
Local Arm (S)                      & -300  &  200  &       &  -5.0 & -1.0  \\
Local Arm (N)                      & -300  &  200  &       &  -6.0 &  1.5  \\
Sagittarius Arm (S)                & -880  &  600  &       &  -3.0 &  9.0  \\
Sagittarius Arm (N)                & -880  &  600  &       &  -5.0 & 10.0  \\
Sagittarius-to-Scutum Interarm (S) & -200  &  400  &       &  -10.0 & 18.0 \\ 
Sagittarius-to-Scutum Interarm (N) & -400  &  400  & 400  &   -10.0 & 18.0  \\ 
Scutum Arm (S)                     & -125  & 1000  &      &   -8.0 & 20.0  \\
Scutum Arm (N)                     & -300  &  630  &      &   -8.0 & 15.0  \\ 
Perseus Arm (S)                    & -500  &  500  &      &   -20.0 & 20.0  \\  
Perseus Arm (N)                    & -500  &  500  & 300  & -20.0 &  8.0  \\  
\enddata
\end{deluxetable*}

\clearpage

\startlongtable
\begin{deluxetable*}{lrrccccc}
\tablecolumns{8}
\tablecaption{Pulsar LOS rejected as outliers from  zonal magnetic field fits 
\label{table:outliers}}
\tablehead{
\colhead{PSR }  &\colhead{$l$} & \colhead{$b$}  & \colhead{\hspace{0cm}Distance}  & \colhead{\hspace{0cm} $\delta\rm RM_{zone}/\cos\theta$\tablenotemark{a} } & \colhead{\hspace{0cm}$\delta\rm DM_{zone}$\tablenotemark{a} } & \colhead{\hspace{0in}  $B$\tablenotemark{a} } &  \colhead{Rejection} \\
[-0cm] \colhead{} & \colhead{ ($^{\rm o}$) } &  \colhead{ ($^{\rm o}$) } & \colhead{\hspace{0cm}(kpc) }  & \colhead{\hspace{0cm} (rad m$^{-2}$) } & \colhead{\hspace{0cm} (pc cm$^{-3}$) } & 
\colhead{ ($\mu$G)}  & \colhead{\hspace{-0in}Criterion\tablenotemark{b}} 
}


\startdata
\cutinhead{Local Arm} 
B0833--45     &  263.6 &   -2.8 &    0.3 &         -31.4 &          68.0 &          -0.6 &   4 \\
B0923--58     &  278.4 &   -5.6 &    0.1 &          47.4 &          57.4 &           1.0 &   4 \\
B0940--55     &  278.6 &   -2.2 &    0.3 &          65.1 &         180.2 &           0.4 &   4 \\
B0959--54     &  280.2 &    0.1 &    0.3 &        -315.3 &         130.3 &          -3.0 &   2 \\
B2022+50     &   86.9 &    7.5 &    2.1 &          46.2 &          33.0 &           1.7 &   4 \\
J2032+4127   &   80.2 &    1.0 &    1.3 &         215.8 &         114.7 &           2.3 &   2,4 \\
\cutinhead{Sagittarius Arm}
B1749--28     &    1.5 &   -1.0 &    0.2 &         468.5 &          48.1 &          12.0 &   4 \\
J1804--2717   &    3.5 &   -2.7 &    0.8 &         173.5 &          22.4 &           9.6 &   4 \\
B1820--31     &    2.1 &   -8.3 &    1.6 &         401.9 &          48.0 &          10.3 &   4 \\
J1920+1110   &   46.2 &   -1.2 &    6.1 &         660.2 &         177.6 &           4.6 &   2 \\
J2017+2043   &   61.4 &   -8.3 &    4.2 &        -146.9 &          52.0 &          -3.5 &   4 \\
J2015+2524   &   65.0 &   -5.3 &    0.9 &          -8.1 &           1.5 &          -6.9 &   4 \\
J2016+1948   &   60.5 &   -8.7 &    2.2 &        -108.1 &          25.8 &          -5.2 &   4 \\
J1730--2304   &    3.1 &    6.0 &    0.6 &         -34.2 &           7.3 &          -5.8 &   4 \\
J1802--2124   &    8.4 &    0.6 &    0.8 &         854.3 &         147.3 &           7.2 &   2 \\
B1822--09     &   21.4 &    1.3 &    0.3 &         131.6 &           9.5 &          17.1 &   4 \\
B1907+12     &   46.2 &    1.6 &    8.1 &        1001.3 &         254.2 &           4.9 &   2 \\
B1911+13     &   47.9 &    1.6 &    5.3 &         612.3 &         140.4 &           5.4 &   2 \\
B1921+17     &   51.7 &    1.0 &    4.0 &         614.9 &         137.2 &           5.5 &   2 \\
\cutinhead{Sagittarius-to-Scutum Interarm} 
J1759--2922   &    1.2 &   -2.9 &    2.4 &         849.6 &          23.7 &          44.3 &   2,4 \\
B1853+01     &   34.6 &   -0.5 &    3.3 &        -217.3 &          15.4 &         -17.4 &   2,4 \\
B1854+00     &   34.4 &   -0.8 &    2.5 &          34.0 &           2.0 &          20.6 &   4 \\
J1901+0254   &   36.6 &   -0.9 &    4.3 &        -206.4 &          99.8 &          -2.6 &   2 \\
B1859+03     &   37.2 &   -0.6 &    7.0 &        -315.6 &         314.6 &          -1.2 &   2 \\
J1907+0740   &   41.6 &   -0.1 &    5.8 &         487.1 &         225.1 &           2.7 &   2 \\
B1907+03     &   38.6 &   -2.3 &    2.9 &        -214.6 &           1.1 &        -245.3 &   2,4 \\
J1910+0728   &   41.7 &   -0.8 &    6.2 &         473.0 &         178.7 &           3.3 &   2 \\
J1915+0752   &   42.6 &   -1.6 &    3.6 &         122.6 &           4.5 &          33.4 &   4 \\
J1832--0836   &   23.1 &    0.3 &    2.5 &         -22.6 &         -46.4 &        --- &   1 \\
J1901+0510   &   38.7 &    0.0 &    5.9 &        1212.0 &         335.9 &           4.5 &   2 \\
J1901+0621   &   39.7 &    0.8 &    2.9 &        -336.9 &           0.5 &        -831.7 &   4 \\
B1900+06     &   39.8 &    0.3 &    7.0 &         463.8 &         406.5 &           1.4 &   2,3 \\
J1902+0723   &   40.7 &    1.0 &    3.3 &        -357.3 &           9.2 &         -48.2 &   4 \\
B1907+10     &   44.8 &    1.0 &    4.8 &         429.5 &          21.2 &          25.0 &   2,4 \\
J1848+0826   &   40.2 &    4.4 &    3.9 &         216.6 &          12.8 &          21.0 &   4 \\
\cutinhead{Scutum Arm}
B1758--29     &    1.4 &   -3.2 &    3.0 &       -378.3 &          41.3 &         -11.3 &   2,4 \\
B1842--04     &   28.2 &   -0.8 &    4.1 &        -379.0 &          73.1 &          -6.4 &   2 \\
J1849--0614   &   27.2 &   -2.5 &    3.5 &         394.5 &           7.6 &          64.3 &   4 \\
J1900--0051   &   33.2 &   -2.5 &    4.2 &         193.7 &           4.7 &          50.7 &   4 \\
J1902--0340   &   31.0 &   -4.2 &    4.0 &         228.0 &          10.2 &          27.5 &   4 \\
B1911--04     &   31.3 &   -7.1 &    4.0 &         -24.1 &           3.5 &          -8.5 &   4 \\
B1917+00     &   36.5 &   -6.2 &    5.9 &          84.1 &         -18.2 &        --- &   1 \\
J1858+0241   &   36.2 &   -0.4 &    5.2 &        -200.3 &          16.2 &         -15.3 &   2,4 \\
J1741--2733   &    0.6 &    1.6 &    3.1 &        -480.0 &          57.1 &         -10.4 &   2,4 \\
B1804--12     &   17.1 &    4.4 &    3.0 &         342.9 &           7.4 &          57.2 &   4 \\
J1829+0000   &   30.5 &    4.8 &    4.3 &        -125.0 &          17.8 &          -8.7 &   4 \\
B1834--06     &   25.2 &    0.0 &    4.1 &        -386.4 &         159.4 &          -3.0 &   2 \\
J1841--0345   &   28.4 &    0.4 &    3.8 &         470.4 &          31.3 &          18.6 &   4 \\
B1839--04     &   28.3 &    0.2 &    3.7 &         326.2 &          26.2 &          15.3 &   4 \\
J1842--0415   &   28.1 &    0.1 &    3.6 &        -168.5 &          18.8 &         -11.1 &   4 \\
B1845--01     &   31.3 &    0.0 &    4.4 &         500.0 &         -39.8 &        ---  &   1 \\
J1849+0127   &   34.0 &    1.0 &    4.7 &        -216.3 &          32.6 &          -8.2 &   4 \\
J1850--0006   &   32.8 &    0.1 &    5.6 &         631.9 &         351.9 &           2.2 &   2 \\
 \cutinhead{Perseus Arm}   
B0329+54     &  145.0 &   -1.2 &    1.7 &        -130.7 &           6.5 &         -24.7 &   4 \\
B0525+21     &  183.9 &   -6.9 &    1.2 &         264.6 &           9.4 &          34.8 &   4 \\ 
B0531+21     &  184.6 &   -5.8 &    2.0 &         292.0 &          14.4 &          25.0 &   4 \\ 
J0538+2817   &  179.7 &   -1.7 &    1.3 &        -164.0 &           0.4 &        -470.9 &   4,5 \\
B2035+36     &   76.7 &   -2.8 &    4.9 &         448.7 &          19.2 &          28.9 &   4 \\
J0215+6218   &  132.6 &    1.0 &    2.0 &         735.8 &          63.2 &          14.4 &   2,4 \\
J0248+6021   &  136.9 &    0.7 &    2.0 &        -307.3 &         349.8 &          -1.1 &   3 \\
B0355+54     &  148.2 &    0.8 &    1.0 &         268.6 &          36.5 &           9.1 &   4 \\
B0450+55     &  152.6 &    7.5 &    1.2 &          47.2 &          -6.5 &           0.0 &   1 \\
B0458+46     &  160.4 &    3.1 &    1.3 &       -1267.2 &          17.4 &         -89.8 &   2,4,5 \\
J0611+30     &  181.6 &    5.5 &    1.1 &         -31.3 &           4.4 &          -8.7 &   5 \\
J0709+0458   &  210.5 &    6.2 &    1.2 &         -25.2 &           0.7 &         -44.0 &   4 \\
J0540+3207   &  176.7 &    0.8 &    1.4 &         -74.8 &          25.4 &          -3.6 &   5 \\
J0711+0931   &  206.7 &    8.8 &    1.2 &         -63.0 &           1.1 &         -73.5 &   4 \\
\enddata
\tablenotetext{a} { The given quantity refers to the value associated with the specified Arm or Interarm Region.}
\tablenotetext{b} { Rejection Criteria: 1:\ negative $\delta$\rm{DM};\ 2\! :\ $\delta${\rm (RM}/$\cos \theta)$\ outlier;\ 3\! :\ $\delta$\rm{DM}\ \rm{outlier};\ 4\! :\  \it{B}\ \rm{outlier}; \\ 5\! :\ ${\bf{B}} \perp~d{\bf{s}} \ ( i.e., |\cos \theta\ | < 0.2).  $}   
\end{deluxetable*}

\begin{acknowledgements}
The authors gratefully acknowledge financial support from the US National Science Foundation, under Grants  AST-1312843 (AC and JW) and 18-14397 (JR).  The Arecibo Observatory was operated by the University of Central Florida under a cooperative agreement with the US National Science Foundation, and in alliance with Yang Enterprises and the Ana G. M\'endez-Universidad Metropolitana. A.P.C is a Vanier Canada Graduate Scholar. 
\end{acknowledgements}

\facility{Arecibo}

\restartappendixnumbering 
\appendix
\section{Pulsars whose LOS intersect the Gum Nebula}

Table A1 
below lists those pulsars in our sample whose
LOS intersect the Gum Nebula.  As noted in \S\ref{sssec:local},
this Nebula has a complicated and poorly known magnetic structure, so
we exclude these pulsars from our fits.
\begin{deluxetable}{lrrrr}
\tablecolumns{5}
\tabletypesize{\scriptsize}
\tablecaption{Pulsars in our Sample Lying Within or Beyond the Gum Nebula}
\label{table:HitGum}
\tabletypesize{\scriptsize}
\tablehead{ 
\colhead{PSR} & \colhead{$l$         } & \colhead{ $b$         }   & \colhead{DM}              & \colhead{Distance}   \\[-8pt]
\colhead{   }  & \colhead{($^{\rm o}$)} & \colhead{($^{\rm o}$)}   & \colhead{(pc cm$^{-3}$) } &  \colhead{(kpc)}     
}
\startdata
J0737-3039A/B&  245.2 &   -4.5 &   48.9 &    1.1 \\
B0736-40     &  254.2 &   -9.2 &  160.9 &    1.6 \\
B0740-28     &  243.8 &   -2.4 &   73.7 &    2.0 \\
J0749-4247   &  257.1 &   -8.3 &  104.6 &    0.6 \\
B0808-47     &  263.3 &   -8.0 &  228.3 &    6.5 \\
J0818-3232   &  251.4 &    1.9 &  131.8 &    0.5 \\
J0820-3826   &  256.5 &   -1.0 &  195.6 &    4.1 \\
J0820-3921   &  257.3 &   -1.6 &  179.4 &    3.6 \\
B0818-41     &  258.7 &   -2.7 &  113.4 &    0.6 \\
J0821-4221   &  259.8 &   -3.1 &  270.6 &    5.8 \\
B0826-34     &  254.0 &    2.6 &   52.2 &    0.4 \\
J0831-4406   &  262.3 &   -2.7 &  254.0 &    5.9 \\
J0834-4159   &  260.9 &   -1.0 &  240.5 &    5.5 \\
J0835-3707   &  257.1 &    2.0 &  112.3 &    0.6 \\
B0835-41     &  260.9 &   -0.3 &  147.3 &    1.5 \\
J0838-2621   &  248.8 &    9.0 &  116.9 &    4.1 \\
B0839-53     &  270.8 &   -7.1 &  156.5 &    0.6 \\
B0840-48     &  267.2 &   -4.1 &  196.8 &    3.1 \\
J0843-5022   &  268.5 &   -4.9 &  178.5 &    1.6 \\
B0844-35     &  257.2 &    4.7 &   94.2 &    0.5 \\
B0853-33     &  256.8 &    7.5 &   86.6 &    0.5 \\
J0855-4644   &  267.0 &   -1.0 &  236.4 &    5.6 \\
J0855-4658   &  267.1 &   -1.2 &  472.7 &   13.7 \\
J0857-4424   &  265.5 &    0.8 &  184.4 &    2.8 \\
J0900-3144   &  256.2 &    9.5 &   75.7 &    0.9 \\
J0901-4624   &  267.4 &   -0.0 &  199.3 &    3.0 \\
B0903-42     &  265.1 &    2.9 &  145.8 &    0.7 \\
J0905-4536   &  267.2 &    1.0 &  179.7 &    2.0 \\
J0905-5127   &  271.6 &   -2.9 &  196.4 &    1.3 \\
J0905-6019   &  278.2 &   -8.8 &   91.4 &    0.4 \\
B0905-51     &  272.2 &   -3.0 &  103.7 &    0.3 \\
B0906-49     &  270.3 &   -1.0 &  180.4 &    1.0 \\
J0912-3851   &  263.2 &    6.6 &   71.5 &    0.3 \\
J0922-4949   &  272.2 &    0.2 &  237.1 &    2.7 \\
B0922-52     &  274.7 &   -1.9 &  152.9 &    0.5 \\
B0932-52     &  275.7 &   -0.7 &  100.0 &    0.3 \\
J0940-5428   &  277.5 &   -1.3 &  134.6 &    0.4 \\
J0941-5244   &  276.4 &    0.1 &  157.9 &    0.4 \\
B0941-56     &  279.3 &   -3.0 &  159.7 &    0.4 \\
J0945-4833   &  274.2 &    3.7 &   98.1 &    0.4 \\
J0954-5430   &  279.0 &   -0.1 &  201.6 &    0.4 \\
B0953-52     &  278.3 &    1.2 &  156.9 &    0.4 \\
J0957-5432   &  279.4 &    0.2 &  226.1 &    0.4 \\
B0957-47     &  275.7 &    5.4 &   92.7 &    0.4 \\
J1001-5559   &  280.7 &   -0.6 &  159.3 &    0.4 \\
B1001-47     &  276.0 &    6.1 &   98.5 &    0.4 \\
B1011-58     &  283.7 &   -2.1 &  383.9 &    3.2 \\
J1013-5934   &  284.1 &   -2.6 &  379.8 &    3.1 \\
J1015-5719   &  283.1 &   -0.6 &  278.1 &    2.7 \\
J1016-5819   &  283.7 &   -1.4 &  252.2 &    2.6 \\
J1016-5857   &  284.1 &   -1.9 &  394.5 &    3.2 \\
B1015-56     &  282.7 &    0.3 &  438.7 &    3.5 \\
J1019-5749   &  283.8 &   -0.7 & 1040.0 &   10.9 \\
J1020-6026   &  285.3 &   -2.8 &  441.5 &    3.3 \\
J1036-4926   &  281.5 &    7.7 &  136.5 &    2.3 \\
\enddata 

\label{table:GumHits}
\end{deluxetable}


\end{document}

%% file: auth.tex
\author[0000-0002-8376-1563]{Alice P.~Curtin}
  \affiliation{Department of Physics, McGill University, 3600 rue University, Montr\'eal, QC H3A 2T8, Canada}
  \affiliation{Trottier Space Institute, McGill University, 3550 rue University, Montr\'eal, QC H3A 2A7, Canada}
  \affiliation{Department of Physics and Astronomy, Carleton College, Northfield, MN USA 55057}

\author[0000-0001-9096-6543]{Joel M. ~Weisberg}
  \affiliation{Department of Physics and Astronomy, Carleton College, Northfield, MN USA 55057}

\author[0000-0002-8923-6065]{Joanna M. ~Rankin}
  \affiliation{Department of Physics, University of Vermont, Burlington, VT  05405 USA}

%% file: galMagRM.bbl
\begin{thebibliography}

\bibitem[Beck(2016)]{Beck2016} Beck, R.\ 2016, \aapr, 24, 4. doi:10.1007/s00159-015-0084-4

\bibitem[Bilitza(2003)]{B2003} Bilitza, D.\ 2003,  Adv. Space Res. 31,  757

\bibitem[Brown et al.(2003)]{Brown2003} Brown, J.~C., Taylor, A.~R., Wielebinski, R., Mueller, P. \ 2003, \apjl, 592, L29


\bibitem[Deller et al.(2019)]{Delleret2019} Deller, A.~T., Goss, W.~M., Brisken, W.~F., et al.\ 2019, \apj, 875, 100

\bibitem[Dickey et al. (2022)]{Dickeyet2022}Dickey, J.~M., West, J., Thomson, A.~J.~M., et al.\ 2022, \apj, 940, 75. doi:10.3847/1538-4357/ac94ce

\bibitem[Efron \& Tibshirani(1991)]{Efron1991} Efron, B. \& Tibshirani, R.\ 1991, Science, 253, 390

\bibitem[Griv et al.(2017)]{Grivet2017} Griv, E., Jiang, I.-G., \& Hou, L.-G.\ 2017, \apj, 844, 118

\bibitem[Han(2017)]{Han2017} Han, J.~L.\ 2017, \araa, 55, 111. doi:10.1146/annurev-astro-091916-055221

\bibitem[Han et al.(2018)]{Han2018} Han, J.~L., Manchester, R.~N., van Straten, W., \& Demorest, P.\ 2018, \apjs, 234, 11 

\bibitem[Heald et al.(2009)]{Heald2009} Heald, G., Braun, R., \& Edmonds, R.\ 2009, \aap, 503, 409. doi:10.1051/0004-6361/200912240

\bibitem[HI4PI Collaboration et al.(2016)]{HI4PIpaper} HI4PI Collaboration, Ben Bekhti, N., Flöer, L., et al.\ 2016, \aap, 594, A116. doi:10.1051/0004-6361/201629178

\bibitem[Hou \& Han(2014)]{HouHan2014} Hou, L.~G., \& Han, J.~L.\ 2014, \aap, 569, A125 

\bibitem[Jansson \& Farrar(2012)]{Jansson2012} Jansson, R. \& Farrar, G.~R.\ 2012, \apj, 757, 14. doi:10.1088/0004-637X/757/1/14


\bibitem[Johnston et al.(2021)]{Johnston2021} Johnston, S., Sobey, C., Dai, S., et al.\ 2021, \mnras, 502, 1253

\bibitem[Khoperskov \& Khrapov(2018)]{Khop2018} Khoperskov, S.~A., \& Khrapov, S.~S.\ 2018, \aap, 609, A104 

\bibitem[Kirichenko et al.(2015)]{Kiriet2015} Kirichenko, A., Danilenko, A., Shternin, P., et al.\ 2015, \apj, 802, 17 

\bibitem[Krause(2019)]{Krause2019} Krause, M.\ 2019, Galaxies, 7, 54. doi:10.3390/galaxies7020054


\bibitem[Kronberg (2016)]{Kronberg2016} Kronberg, P.P. (2016) ‘Magnetic field configurations in large galaxies’, in Cosmic Magnetic Fields. Cambridge: Cambridge University Press (Cambridge Astrophysics), pp. 72–99.

\bibitem[K{\"u}t{\"u}kc{\"u} et al.(2022)]{Kutkcu2022} K{\"u}t{\"u}kc{\"u}, P., Ankay, A., Yazgan, E., et al.\ 2022, \mnras, 511, 4669. doi:10.1093/mnras/stac346

\bibitem[L{\'e}pine et al.(2017)]{Lepineet2017} L{\'e}pine, J.~R.~D., Michtchenko, T.~A., Barros, D.~A., \& Vieira, R.~S.~S.\ 2017,
 \apj, 843, 48 

\bibitem[Lorimer \& Kramer(2012)]{LK2012} Lorimer, D.~R., \& Kramer, M.\ 2012, Handbook of 
Pulsar Astronomy, Cambridge, UK: Cambridge University Press. 

\bibitem[Ma et al.(2020)]{Maet2020} Ma, Y.~K., Mao, S.~A., Ordog, A., et al.\ 2020, \mnras, 497, 3097   doi:10.1093/mnras/staa2105

\bibitem[Manchester(1972)]{Manchester1972} Manchester, R.~N.\ 1972, \apj, 172, 43. doi:10.1086/151326

\bibitem[Manchester et al.(2005)]{Manch2005} Manchester, R.~N., Hobbs, G.~B., Teoh, A., \& Hobbs, M.\ 2005, \aj, 129, 1993 

\bibitem[Men et al.(2008)]{Men2008} Men, H., Ferri{\`e}re, K., \& Han, J.~L.\ 2008, \aap, 486, 819. doi:10.1051/0004-6361:20078683

\bibitem[Mitra et al.(2003)]{Mitra2003} Mitra, D., Wielebinski, R., Kramer, M., et al.\ 2003, \aap, 398, 993. doi:10.1051/0004-6361:20021702

\bibitem[Ocker et al.(2024)]{Ocker2024} Ocker, S.~K., Anderson, L.~D., Lazio, T.~J.~W., et al.\ 2024, arXiv:2406.07664. doi:10.48550/arXiv.2406.07664

\bibitem[Ohno \& Shibata(1993)]{Ohno1993} Ohno, H. \& Shibata, S.\ 1993, \mnras, 262, 953

\bibitem[Ordog et al.(2017)]{Ordoget2017} Ordog, A., Brown, J.~C., Kothes, R., et al.\ 2017, \aap, 603, A15. doi:10.1051/0004-6361/201730740

\bibitem[Purcell et al.(2015)]{Purcellet2015} Purcell, C.~R., Gaensler, B.~M., Sun, X.~H., et al.\ 2015, \apj, 804, 22 

\bibitem[Rankin et al.(2023)]{Rankin2023} Rankin, J., Venkataraman, A., Weisberg, J. M., \& Curtin, A. P.\ 2023, \mnras, 524, 5042

\bibitem[Reid et al.(2016)]{Reidet2016} Reid, M.~J., Dame, T.~M., Menten, K.~M., \& Brunthaler, A.\ 2016, \apj, 823, 77 


\bibitem[Reid et al.(2019)]{Reidet2019} Reid, M.~J., Menten, K.~M., Brunthaler, A., et al.\ 2019, \apj, 885, 131 

\bibitem[Shanahan et al.(2019)]{Shanahanet2019} Shanahan, R., Lemmer, S.~J., Stil, J.~M., et al.\ 2019, \apjl, 887, L7 
(S19) 

\bibitem[Sobey et al.(2019)]{Sobeyet2019} Sobey, C., Bilous, A.~V., Grie{\ss}meier, J.-M., et al.\ 2019, \mnras, 484, 3646 

\bibitem[Sun et al.(2008)]{Sunet2008} Sun, X.~H., Reich, W., Waelkens, A., et al.\ 2008, \aap, 477, 573. doi:10.1051/0004-6361:20078671

\bibitem[Unger \& Farrar(2023)]{Unger2023} Unger, M. \& Farrar, G.~R.\ 2023, arXiv:2311.12120. doi:10.48550/arXiv.2311.12120

\bibitem[Valleé (2008)]{Vallee2008} Vall{\'e}e, J.~P.\ 2008, \apj, 681, 303. doi:10.1086/588577

\bibitem[Van Eck et al.(2011)]{VanEck2011} Van Eck, C.~L., Brown, J.~C., Stil, J.~M., et al.\ 2011, \apj, 728, 97

\bibitem[Weisberg et al.(2004)]{Wetal2004} Weisberg, J.~M., Cordes, J.~M., Kuan, B.,
Devine, K.~E., Green, J.~T., \& Backer, D.~C.\ 2004, \apjs, 150, 317 

\bibitem[Xu et al.(2013)]{Xuet2013} Xu, Y., Li, J.~J., Reid, M.~J., et al.\ 2013, \apj, 769, 15 

\bibitem[Xu et al.(2018)]{Xuet2018} Xu, Y., Bian, S.~B., Reid, M.~J., et al.\ 2018, \aap, 616, L15

\bibitem[Xu \& Han(2019)]{XuHan2019} Xu, J. \& Han, J.~L.\ 2019, \mnras, 486, 4275 

\bibitem[Xu et al.(2021)]{Xuet2021} Xu, Y., Hou, L.~G., Bian, S.~B., et al.\ 2021, \aap, 645, L8

\bibitem[Xu et al.(2022)]{Xuet2022} Xu, J., Han, J., Wang, P., Yan, Yi. \ 2022, Science China Physics, Mechanics, and Astronomy, 65, 129704. doi:10.1007/s11433-022-2033-2

\bibitem[Xu \& Han(2024)]{XuHan2024} Xu, J. \& Han, J.~L.\ 2024, \apj, 966, 240. doi:10.3847/1538-4357/ad3a61


\bibitem[Yao, Manchester, \& Wang(2017)]{YMW2017} Yao, J.~M., Manchester, R.~N., \& Wang, N.\ 2017, \apj, 
835, 29 (YMW17)

\end{thebibliography}
